\documentclass[letterpaper,twocolumn,10pt]{article}
\usepackage{usenix}
    
 \usepackage{amsmath}
 \usepackage{xspace}
 \usepackage{cleveref}
 \usepackage{graphicx}
\usepackage{subcaption}
\usepackage{caption}
\usepackage{tikz}
\usepackage{booktabs}
\usepackage{multirow}
\usepackage{array}
\usepackage{enumerate}
\usepackage{enumitem}
\usepackage{makecell}

\usepackage[binary-units=true]{siunitx}
\newcommand{\SIx}[1]{\num{#1}\relax}
\DeclareSIUnit\per{/}
\DeclareSIUnit\dollar{\$}
\DeclareSIUnit{\month}{month}
\DeclareSIUnit{\thousand}{k}
\DeclareSIUnit{\million}{M}

\usepackage{array}
\newcolumntype{L}[1]{>{\raggedright\let\newline\\\arraybackslash\hspace{0pt}}m{#1}}
\newcolumntype{C}[1]{>{\centering\let\newline\\\arraybackslash\hspace{0pt}}m{#1}}
\newcolumntype{R}[1]{>{\raggedleft\let\newline\\\arraybackslash\hspace{0pt}}m{#1}}

\usepackage[skins]{tcolorbox}

\newcounter{primitivecounter}
\crefname{primitivecounter}{Attack Primitive}{Attack Primitives}

\newtcolorbox[use counter=primitivecounter]{primitivebox}[2]{
  enhanced, width=\hsize,left=1pt,right=1pt,top=1pt,bottom=1pt,colback=red!6!white,boxrule=1pt,colframe=red!30!white,
  label type=primitivecounter,
  label = #1,
  title = \textcolor{black}{\bf Attack Primitive~\arabic{primitivecounter}~#2}
}

\usepackage{pgfplots}
\usepgfplotslibrary{fillbetween}
\usepackage{tikz}
\usetikzlibrary{matrix,positioning}
\usetikzlibrary{external}
\usepackage{pgfplots}
\usepackage{pgfplotstable}
\usetikzlibrary{pgfplots.groupplots}
\usetikzlibrary{arrows}
\usetikzlibrary{patterns}
\usetikzlibrary{positioning}
\usetikzlibrary{decorations.pathreplacing}
\usetikzlibrary{shapes.arrows}
\usetikzlibrary{shapes.geometric,shapes.misc}
\usetikzlibrary{pgfplots.groupplots}
\pgfplotsset{compat=newest}
\pgfkeys{/pgf/number format/.cd,1000 sep={}}

\pgfplotsset{
    discard if/.style 2 args={
        x filter/.code={
            \edef\tempa{\thisrow{#1}}
            \edef\tempb{#2}
            \ifx\tempa\tempb
                
            \fi
        }
    },
    discard if not/.style 2 args={
        x filter/.code={
            \edef\tempa{\thisrow{#1}}
            \edef\tempb{#2}
            \ifx\tempa\tempb
            \else
                
            \fi
        }
    }
}

\pgfdeclareshape{rectangle with diagonal fill}
{
    \inheritsavedanchors[from=rectangle]
    \inheritanchorborder[from=rectangle]
    \inheritanchor[from=rectangle]{north}
    \inheritanchor[from=rectangle]{north west}
    \inheritanchor[from=rectangle]{north east}
    \inheritanchor[from=rectangle]{center}
    \inheritanchor[from=rectangle]{west}
    \inheritanchor[from=rectangle]{east}
    \inheritanchor[from=rectangle]{mid}
    \inheritanchor[from=rectangle]{mid west}
    \inheritanchor[from=rectangle]{mid east}
    \inheritanchor[from=rectangle]{base}
    \inheritanchor[from=rectangle]{base west}
    \inheritanchor[from=rectangle]{base east}
    \inheritanchor[from=rectangle]{south}
    \inheritanchor[from=rectangle]{south west}
    \inheritanchor[from=rectangle]{south east}

    \inheritbackgroundpath[from=rectangle]
    \inheritbeforebackgroundpath[from=rectangle]
    \inheritbehindforegroundpath[from=rectangle]
    \inheritforegroundpath[from=rectangle]
    \inheritbeforeforegroundpath[from=rectangle]

    \behindbackgroundpath{%
        \pgfextractx{\pgf@xa}{\southwest}%
        \pgfextracty{\pgf@ya}{\southwest}%
        \pgfextractx{\pgf@xb}{\northeast}%
        \pgfextracty{\pgf@yb}{\northeast}%
        \ifpgf@diagonal@lefttoright
            \def\pgf@diagonal@point@a{\pgfpoint{\pgf@xa}{\pgf@yb}}%
            \def\pgf@diagonal@point@b{\pgfpoint{\pgf@xb}{\pgf@ya}}%
        \else
            \def\pgf@diagonal@point@a{\southwest}%
            \def\pgf@diagonal@point@b{\northeast}%
        \fi
        \pgfpathmoveto{\pgf@diagonal@point@a}%
        \pgfpathlineto{\northeast}%
        \pgfpathlineto{\pgfpoint{\pgf@xb}{\pgf@ya}}%
        \pgfpathclose
        \ifpgf@diagonal@lefttoright
            \color{\pgf@diagonal@top@color}%
        \else
            \color{\pgf@diagonal@bottom@color}%
        \fi
        \pgfusepath{fill}%
        \pgfpathmoveto{\pgfpoint{\pgf@xa}{\pgf@yb}}%
        \pgfpathlineto{\southwest}%
        \pgfpathlineto{\pgf@diagonal@point@b}%
        \pgfpathclose
        \ifpgf@diagonal@lefttoright
            \color{\pgf@diagonal@bottom@color}%
        \else
            \color{\pgf@diagonal@top@color}%
        \fi
        \pgfusepath{fill}%
    }
}

\makeatletter
\newcommand\resetstackedplots{%
\pgfplots@stacked@isfirstplottrue
}
\makeatother

\newcommand{\eg}{e.g.,\xspace}
\newcommand{\ie}{i.e.,\xspace}
\newcommand*{\stealing}{{CacheExposer\xspace}}
\newcommand*{\poison}{{CachePoison\xspace}}
\newcommand*{\diffusiondb}{{DiffusionDB\xspace}}
\newcommand*{\lexica}{{Lexica\xspace}}

\newcommand*\circled[1]{\tikz[baseline=(char.base)]{
            \node[circle,fill=.,inner sep=0.8pt] (char) {\textcolor{white}{#1}};}}

\begin{document}

\title{Attacks on Approximate Caches in Text-to-Image Diffusion Models}

\author{
{\rm Desen Sun}\\
University of Waterloo
\and
{\rm Shuncheng Jie}\\
University of Waterloo
\and
{\rm Sihang Liu}\\
University of Waterloo
}

\maketitle
\begin{abstract}
Diffusion models are a powerful class of generative models that produce images and other content from user prompts, but they are computationally intensive. 
To mitigate this cost, recent academic and industry work has adopted approximate caching, which reuses intermediate states from similar prompts in a cache.
While efficient, this optimization introduces new security risks by breaking isolation among users.
This paper provides a comprehensive assessment of the security vulnerabilities introduced by approximate caching.
First, we demonstrate a remote covert channel established with the approximate cache, where a sender injects prompts with special keywords into the cache system and a receiver can recover that even after days, to exchange information. 
Second, we introduce a prompt stealing attack using the approximate cache, where an attacker can recover existing cached prompts from hits.
Finally, we introduce a poisoning attack that embeds the attacker's logos into the previously stolen prompt, leading to unexpected logo rendering for the requests that hit the poisoned cache prompts. 
These attacks are all performed remotely through the serving system, demonstrating severe security vulnerabilities in approximate caching. The code for this work is available.\footnote{\url{https://doi.org/10.5281/zenodo.17957900}}

\end{abstract}


\section{Introduction}
Over recent years, generative models have developed rapidly.
Among them, text-to-image diffusion models dominate image generation in commercial systems, such as those from Google \cite{Saharia_NIPS_2022_Photorealistic}, Adobe \cite{firely} and OpenAI \cite{dalle3}. Thanks to their remarkable ability in generating high-quality images, users leverage them for scene and character designs \cite{Sun_2025_CVPR_DRiVE,Wang_ACM_MM_2024_Evolving,Chen_2025_CVPR_POSTA,Gao_2025_CVPR_PosterMaker,Ma_Deng_Chen_Du_Lu_Yang_2025,Bokhovkin_2025_CVPR,Huang_2025_CVPR_MIDI,Yang_2025_CVPR_Prometheus,eldesokey2025buildascene}, artwork construction \cite{Wang_survey_2025_Diffusion}, and advertisement \cite{DU_ECCV_2024_Towards}.
Users only need to provide their prompts that describe the task, and the diffusion model can generate the image as specified. 

Although diffusion models are powerful, they suffer from a high computational overhead.
It takes seconds to generate a single image.
Such overhead mainly comes from the computation pattern, where a sequence of denoising steps are applied to an initial noise to convert it to the final image. 
Therefore, reducing the number of steps is a key approach to saving computation. 
Caching is a promising technique that has attracted attention from both academia and industry \cite{nirvana, xia2025modmefficientservingimage,fast_stable_diffusion,gpt_cache}. 
For instance, Pinecone \cite{fast_stable_diffusion} and Zilliz~\cite{gpt_cache} use semantic caches to reuse existing results if a previous prompt aligns with the current one. 
Approximate cache is another type of optimization \cite{nirvana,xia2025modmefficientservingimage}. 
Instead of directly reusing the prior generations, it reuses intermediate states from similar prompts in the diffusion execution, and resumes the denoising steps to save computation while achieving better text–image alignment.
For instance, NIRVANA \cite{nirvana} from Adobe is an approximate cache for diffusion models.

With the widespread use of text-to-image diffusion models, adversaries also exploit these models for their malicious purposes. 
Some studies take advantage of the diffusion model as a secret channel, embedding private messages and transmitting the output images \cite{chen2025parasitesteganographybasedbackdoorattack,mahfuz2025psyducktrainingfreesteganographylatent,Pulsar,CRoSS,LDStega}. 
Another type of attack steals high-quality prompts to save the prompt design cost~\cite{xinyue_prompt_stealing,wu-etal-2025-vulnerability,cross_modal_prompt_inversion}. 
There are also attacks that aim to poison the diffusion model and inject specific objects or concepts into output images \cite{ding24ccsunderstanding,wang2024the, guo2025rededitingrelationshipdrivenprecisebackdoor,pan24nipsfrom,huang2025implicitbiasinjectionattacks,Shan_SP_2024_Nightshade,wu2023proactive,nasehbackdooring}.

However, existing attacks focus on the model, while vulnerabilities of the diffusion model serving system remain largely underexplored. 
Approximate cache is promising in improving the efficiency of diffusion models, but it introduces additional components to the image generation system, potentially widening the attack surface.
The goal of this work is to comprehensively assess the security implications of approximate caching and exploit its vulnerabilities.

To analyze approximate cache, we take NIRVANA \cite{nirvana}, the state-of-the-art approximate cache from Adobe by replicating their design, and deploy two widely used text-to-image diffusion models, FLUX \cite{flux2024} and Stable Diffusion 3 (SD3) \cite{sd3}, and two real-world prompt datasets, \diffusiondb{} \cite{diffusiondb} and \lexica{} \cite{xinyue_prompt_stealing}.
The serving system is deployed on cloud GPUs to make the environment realistic.
Similar to conventional caches, like those in CPUs or file systems, approximate cache reduces the image generation latency. 
Our evaluation shows that cache hit or miss latency can be clearly distinguished. 
On an H100 GPU, skipping 10\,\% of steps yields \SI{0.92}{\second} and \SI{0.48}{\second} lower latency for FLUX and SD3, respectively.
This timing difference can be leveraged by an attacker to determine whether their prompt hits or misses the cached prompts, which we refer to as \textit{Attack~Primitive~1}. 

However, an approximate cache system is different from conventional caches.
Unlike a conventional cache, where every access hits a specific entry, whether a prompt hits an entry in an approximate cache depends on prompt similarity, which largely varies by prompts.
Thus, even though an attacker may probe the cache and receive a cache hit, the output cannot be directly used to derive cached states. 
We find that generations hitting the same cached prompt have higher similarity, as they preserve part of the original content, such as structure and layout.  
Therefore, such similarity can be leveraged to determine whether two cache hits originate from the same cached prompt. We refer to this characteristic as \textit{Attack~Primitive~2}.




Based on these attack primitives, we present three attacks.
In these attacks, we assume an attack model where the attacker can only access the serving system like normal users and does not have prior knowledge about any user prompts in the cache. 

First, we demonstrate a remote covert channel, where a sender and a receiver exchange messages through a diffusion model service with approximate caching (\Cref{sec:covert}).
Using special keywords and markers, the sender injects special prompts to the cache.
Afterward, the receiver probes the approximate cache to detect whether the sender's prompts exist.
The receiver uses a combination of generation timing (\Cref{box:primitive_1}) and image content detection (\Cref{box:primitive_2}) to confirm that the cache was injected by the sender. Cache hits and misses are encoded as 1s and 0s.
Using a number of keywords, the sender can transmit a message with a 97.8\,\% accuracy under the FLUX model \cite{flux2024}. 
Using real-world prompt traces from \diffusiondb{}, we demonstrate that sender's prompts stay in the cache for over 44 hours, making this channel stealthy.


Second, we perform a prompt stealing attack, \stealing{} (\Cref{sec:stealing}).
The attacker probes the approximate cache and applies \Cref{box:primitive_1,box:primitive_2}, leveraging both timing and structural similarity of generated images to identify which probing prompts map to the same cached prompt.
These prompts are then used to recover the target prompt.
We evaluate this attack on FLUX and SD3, and two datasets, \diffusiondb{} and \lexica{}, with real-world user prompts.
The results indicate that the attacker's stolen prompts have an average of 0.78 cosine similarity compared to the original ones.

Lastly, we demonstrate a \poison{} attack where an attacker can inject their content (\ie{} logos in this attack) in the diffusion model's cache, and poison other users' generations (\Cref{sec:poison}). 
The attacker first uses an embedding-space logo injection model to seamlessly embed the logo in stolen prompts, ensuring a high hit rate by other users. 
Then, if a user hits the attacker's cached prompt, the output image will include the logo, even though it is not specified in the prompt.

We summarize the contributions as the following:

\begin{itemize}[leftmargin=*,noitemsep,partopsep=0pt,topsep=0pt,parsep=0pt]
    \item To the best of our knowledge, this is the first work that exploits security vulnerabilities in the approximate cache for text-to-image diffusion models. 
    \item We demonstrate a remote covert channel through an approximate caching system, leveraging images generated by special words to secretly exchange information.
    \item We design a prompt stealing attack, \stealing{}. Without requiring users' output images, the attacker can infer user prompts in the approximate cache. 
    \item We introduce a poison attack, \poison{}.
    Without manipulating the model or training dataset, the attacker can poison the output of user prompts with specific logos.
\end{itemize}

\section{Background} \label{sec:background}
In this section, we introduce the architecture of text-to-image diffusion models and a key optimization to diffusion models using approximate caching.

\begin{figure}
    \centering
    \includegraphics[width=1\linewidth]{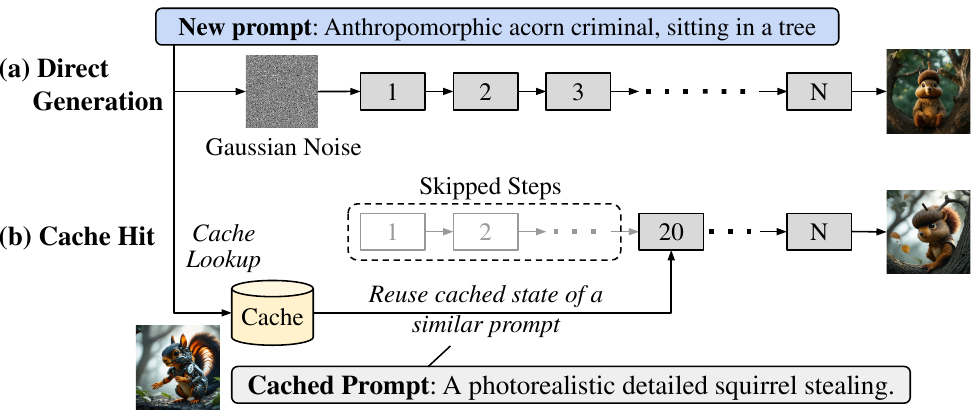}
    \caption{ Diffusion model generation (a) directly from Gaussian noise and (b) from an approximate cache.}
    \label{fig:background}
\end{figure}

\subsection{Text-to-Image Diffusion Models}

Diffusion model is a generative model with iterative computation \cite{ddpm}. 
During inference, a text-to-image diffusion model takes a Gaussian noise image as input, predicts and eliminates noise by a number of steps, and eventually generates a high-quality image.
\Cref{fig:background}a demonstrates how the diffusion model generates an image with a given prompt. 
The diffusion model processes the initial Gaussian noise for multiple denoising steps. In each step, it predicts the potential noise and eliminates such noise from the input. The prompt is used to guide the noise prediction in every step. After a number of denoising steps, the noise will eventually be converted to a high-quality image. 

\subsection{Approximate Cache Serving System}
\label{subsec:approx_cache}

A major challenge of diffusion models is their substantial computational cost, primarily stemming from the iterative denoising process. 
Inspired by semantic caching in large language models (LLMs) \cite{Mohandoss_CAI_2024_Context,li2024scalmsemanticcachingautomated,gill2025meancacheusercentricsemanticcaching,bang-2023-gptcache} and diffusion models \cite{fast_stable_diffusion,gpt_cache,beaumont-2022-clip-retrieval}, which enables reuse of outputs from prior prompts, recent work has introduced approximate caching that reuses intermediate states from previous generations \cite{nirvana,sun2024flexcacheflexibleapproximatecache,xia2025modmefficientservingimage}, achieving a balance between quality and efficiency.
The central insight of approximate caching is that similar prompts yield similar outputs. Thus, the intermediate state from a prior, similar prompt can serve as the starting point for a new prompt, allowing the system to skip already completed denoising steps.
Typically, the number of skipped steps is determined by the \textit{semantic similarity} between the new prompt and the cached prompt, where higher similarity allows the new prompt to skip more steps.
Prior work demonstrates that up to 50\,\% of denoising steps can be reused without noticeable quality degradation~\cite{nirvana}, with every 10\,\% of steps cached to enable flexible reuse across varying similarity levels.

\Cref{fig:background}b illustrates how an approximate cache works. 
When a new prompt arrives, 
the approximate caching system first converts the prompt to an embedding using the Contrastive Language-Image Pre-Training (CLIP) model \cite{pmlr-v139-radford21a} and calculates the cosine similarity between the new prompt and the existing cached prompts. 
If the similarity score exceeds a threshold, the system considers this as a hit and the request will reuse the corresponding cached state.
Prior work has shown that a similarity over 0.65 \cite{nirvana} can be regarded as a hit. 
The new prompt will continue with denoising steps from the middle and skip the prior steps (20 steps in \Cref{fig:background}b).

Moreover, cache management also plays a critical role in the effectiveness of approximate caching.
In Adobe’s NIRVANA system \cite{nirvana}, a prompt is inserted into the cache whenever no existing cache entry is hit. To evict stale entries, NIRVANA employs a Least Computationally Beneficial and Frequently Used (LCBFU) policy, which removes cache items that provide minimal computational savings and are infrequently reused. In contrast, more recent work adopts simpler strategies such as FIFO, ensuring that recently accessed prompts remain available in the cache \cite{xia2025modmefficientservingimage}.

\section{Vulnerabilities of Diffusion Models}

In this section, we discuss the existing attacks on diffusion models and the new security implications due to caching.

\subsection{Attacks on Diffusion Models}

Although powerful in generating content, recent work has shown that diffusion models have security concerns.
One category of attacks notices that the diffusion model's output inherently contains massive information, making it easy for steganography. Previous studies have demonstrated attack techniques that embed secret messages into images \cite{chen2025parasitesteganographybasedbackdoorattack,mahfuz2025psyducktrainingfreesteganographylatent,Pulsar,CRoSS,LDStega}. 
These attacks usually require an attacker-controlled model to manipulate generated images. 
Another category of attacks aims at recovering prompts that generated a certain image, as the engineering of prompts is critical in generation. 
These attacks convert the generated images back to prompts \cite{xinyue_prompt_stealing} by taking publicly available generated images (\eg those posted on the internet) as input and training a model to enable recovery.
%
%
Some attacks also attempt to manipulate the generated content, such as inserting the attacker's brand information or bias to the model to guide the generation process \cite{ding24ccsunderstanding,wang2024the, guo2025rededitingrelationshipdrivenprecisebackdoor,pan24nipsfrom,huang2025implicitbiasinjectionattacks,nasehbackdooring}.
Users will receive images with such an injection. Typically, attackers enable such attacks by polluting the training data \cite{jang2025silentbranding,nasehbackdooring,chen2025parasitesteganographybasedbackdoorattack}.

\subsection{New Vulnerabilities from  Caching }

Although approximate caching is a promising optimization approach, it introduces an additional component to the diffusion model serving system, which stores and reuses prior prompts, potentially extending the attack surface. 
Therefore, the goal of this work is to analyze and assess the security risks associated with approximate cache on diffusion models. 


\subsubsection{Timing of Approximate Cache}
\label{subsubsec:cache_timing}

\begin{figure}[t]
    \centering
    \begin{subfigure}[t]{1\columnwidth}
    \centering
    \begin{tikzpicture} 
    \begin{axis}[%
    width=1\hsize,
    style={font=\footnotesize},
    hide axis,
    xmin=10,
    xmax=50,
    ymin=0,
    ymax=0.4,
    legend columns=7,
    column sep=1ex,
    legend image post style={scale=0.5}, 
    legend style={
        cells={align=center},
        anchor=south,
        at={(0.5,0.5)},
        draw=none,
        fill=none,
        column sep=0.5ex,
    },
    ]
    \addlegendimage{area legend, fill = gray, draw=gray, fill opacity=0.5}
    \addlegendentry{Miss};
    \addlegendimage{empty legend}
    \addlegendentry{Steps skipped: }
    \addlegendimage{area legend, fill = green!60!black, draw=green!60!black, fill opacity=0.5}
    \addlegendentry{10\%};
    \addlegendimage{area legend, fill = purple!60!white, draw=purple!60!white, fill opacity=0.5}
    \addlegendentry{20\%};
    \addlegendimage{area legend, fill = orange!60!white, draw=orange!60!white, fill opacity=0.5}
    \addlegendentry{30\%};
    \addlegendimage{area legend, fill = red!60!white, draw=red!60!white, fill opacity=0.5}
    \addlegendentry{40\%};
    \addlegendimage{area legend, fill = blue!60!white, draw=blue!60!white, fill opacity=0.5}
    \addlegendentry{50\%};

    \end{axis}
\end{tikzpicture}
    \end{subfigure}
    \begin{subfigure}[t]{0.48\columnwidth}
    \centering
        \begin{tikzpicture}
\begin{axis}[
ybar,
style={font=\footnotesize},
xlabel={Latency (s)},
ylabel={Frequency (\%)},
width=1\linewidth,
scaled y ticks=false,
xtick pos=bottom,
ytick pos=left,
xtick style={draw=none},
xmin=8,
xmax=22,
ymin=0,
clip=false,
height=3.3cm,
ymajorgrids=true,
legend cell align=left,
legend columns=3,
legend style={
cells={align=left},
anchor=north,
at={(0.45,1.42)},
draw=none,
fill=none,
column sep=0.5ex,
},
legend image code/.code={%
    \draw[#1,draw] (0cm,-0.04cm) rectangle (0.12cm,0.12cm);
}
]
\addplot+[
    fill = gray, draw=gray, fill opacity=0.5,
    hist={bins=90,data min=8, data max=22},
    y filter/.expression={y*100/100},
] table[col sep=comma,y=miss] {fig/csv/hit_miss_latency_dist/flux_a100.csv};
\addplot+[
    fill = green!60!black, draw=green!60!black, fill opacity=0.5,
    hist={bins=90,data min=8, data max=22},
    y filter/.expression={y*100/100},
] table[col sep=comma,y=10p] {fig/csv/hit_miss_latency_dist/flux_a100.csv};
\addplot+[
    fill = purple!60!white, draw=purple!60!white, fill opacity=0.5,
    hist={bins=90,data min=8, data max=22},
    y filter/.expression={y*100/100},
] table[col sep=comma,y=20p] {fig/csv/hit_miss_latency_dist/flux_a100.csv};
\addplot+[
    fill = orange!60!white, draw=orange!60!white, fill opacity=0.5,
    hist={bins=90,data min=8, data max=22},
    y filter/.expression={y*100/100},
] table[col sep=comma,y=30p] {fig/csv/hit_miss_latency_dist/flux_a100.csv};
\addplot+[
    fill = red!60!white, draw=red!60!white, fill opacity=0.5,
    hist={bins=90,data min=8, data max=22},
    y filter/.expression={y*100/100},
] table[col sep=comma,y=40p] {fig/csv/hit_miss_latency_dist/flux_a100.csv};
\addplot+[
    fill = blue!60!white, draw=blue!60!white, fill opacity=0.5,
    hist={bins=90,data min=8, data max=22},
    y filter/.expression={y*100/100},
] table[col sep=comma,y=50p] {fig/csv/hit_miss_latency_dist/flux_a100.csv};



\end{axis}
\end{tikzpicture}
        \caption{FLUX on A100}
        \label{fig:flux-a100}
    \end{subfigure}
    \hfill
    \begin{subfigure}[t]{0.48\columnwidth}
    \centering
        \begin{tikzpicture}
\begin{axis}[
ybar,
style={font=\footnotesize},
xlabel={Latency (s)},
ylabel={Frequency (\%)},
width=1\linewidth,
scaled y ticks=false,
xtick pos=bottom,
ytick pos=left,    
xtick style={draw=none},
xmin=4,
xmax=11,
ymin=0,
clip=false,
height=3.3cm,
ymajorgrids=true,
legend cell align=left,
legend columns=3,
legend style={
cells={align=left},
anchor=north,
at={(0.45,1.42)},
draw=none,
fill=none,
column sep=0.5ex,
},
legend image code/.code={%
    \draw[#1,draw] (0cm,-0.04cm) rectangle (0.12cm,0.12cm);
}
]
\addplot+[
    fill = gray, draw=gray, fill opacity=0.5,
    hist={bins=90,data min=4, data max=10},
    y filter/.expression={y*100/100},
] table[col sep=comma,y=miss] {fig/csv/hit_miss_latency_dist/sd3_a100.csv};
\addplot+[
    fill = green!60!black, draw=green!60!black, fill opacity=0.5,
    hist={bins=90,data min=4, data max=10},
    y filter/.expression={y*100/100},
] table[col sep=comma,y=10p] {fig/csv/hit_miss_latency_dist/sd3_a100.csv};
\addplot+[
    fill = purple!60!white, draw=purple!60!white, fill opacity=0.5,
    hist={bins=90,data min=4, data max=10},
    y filter/.expression={y*100/100},
] table[col sep=comma,y=20p] {fig/csv/hit_miss_latency_dist/sd3_a100.csv};
\addplot+[
    fill = orange!60!white, draw=orange!60!white, fill opacity=0.5,
    hist={bins=90,data min=4, data max=10},
    y filter/.expression={y*100/100},
] table[col sep=comma,y=30p] {fig/csv/hit_miss_latency_dist/sd3_a100.csv};
\addplot+[
    fill = red!60!white, draw=red!60!white, fill opacity=0.5,
    hist={bins=90,data min=4, data max=10},
    y filter/.expression={y*100/100},
] table[col sep=comma,y=40p] {fig/csv/hit_miss_latency_dist/sd3_a100.csv};
\addplot+[
    fill = blue!60!white, draw=blue!60!white, fill opacity=0.5,
    hist={bins=90,data min=4, data max=10},
    y filter/.expression={y*100/100},
] table[col sep=comma,y=50p] {fig/csv/hit_miss_latency_dist/sd3_a100.csv};



\end{axis}
\end{tikzpicture}
        \caption{SD3 on A100}
        \label{fig:sd3-a100}
    \end{subfigure}

    \begin{subfigure}[t]{0.48\columnwidth}
    \centering
        \begin{tikzpicture}
\begin{axis}[
ybar,
style={font=\footnotesize},
xlabel={Latency (s)},
ylabel={Frequency (\%)},
width=1\linewidth,
scaled y ticks=false,
xtick pos=bottom,
ytick pos=left,
xtick style={draw=none},
xmin=4,
xmax=11,
ymin=0,
clip=false,
height=3.3cm,
ymajorgrids=true,
legend cell align=left,
legend columns=3,
legend style={
cells={align=left},
anchor=north,
at={(0.45,1.42)},
draw=none,
fill=none,
column sep=0.5ex,
},
legend image code/.code={%
    \draw[#1,draw] (0cm,-0.04cm) rectangle (0.12cm,0.12cm);
}
]
\addplot+[
    fill = gray, draw=gray, fill opacity=0.5,
    hist={bins=90,data min=5, data max=11},
    y filter/.expression={y*100/100},
] table[col sep=comma,y=miss] {fig/csv/hit_miss_latency_dist/flux_h100.csv};
\addplot+[
    fill = green!60!black, draw=green!60!black, fill opacity=0.5,
    hist={bins=90,data min=5, data max=11},
    y filter/.expression={y*100/100},
] table[col sep=comma,y=10p] {fig/csv/hit_miss_latency_dist/flux_h100.csv};
\addplot+[
    fill = purple!60!white, draw=purple!60!white, fill opacity=0.5,
    hist={bins=90,data min=5, data max=11},
    y filter/.expression={y*100/100},
] table[col sep=comma,y=20p] {fig/csv/hit_miss_latency_dist/flux_h100.csv};
\addplot+[
    fill = orange!60!white, draw=orange!60!white, fill opacity=0.5,
    hist={bins=90,data min=5, data max=11},
    y filter/.expression={y*100/100},
] table[col sep=comma,y=30p] {fig/csv/hit_miss_latency_dist/flux_h100.csv};
\addplot+[
    fill = red!60!white, draw=red!60!white, fill opacity=0.5,
    hist={bins=90,data min=5, data max=11},
    y filter/.expression={y*100/100},
] table[col sep=comma,y=40p] {fig/csv/hit_miss_latency_dist/flux_h100.csv};
\addplot+[
    fill = blue!60!white, draw=blue!60!white, fill opacity=0.5,
    hist={bins=90,data min=5, data max=11},
    y filter/.expression={y*100/100},
] table[col sep=comma,y=50p] {fig/csv/hit_miss_latency_dist/flux_h100.csv};



\end{axis}
\end{tikzpicture}
        \caption{FLUX on H100}
        \label{fig:flux-h100}
    \end{subfigure}
    \hfill
    \begin{subfigure}[t]{0.48\columnwidth}
    \centering
        \begin{tikzpicture}
\begin{axis}[
ybar,
style={font=\footnotesize},
xlabel={Latency (s)},
ylabel={Frequency (\%)},
width=1\linewidth,
scaled y ticks=false,
xtick pos=bottom,
ytick pos=left,
xtick style={draw=none},
xmin=2.5,
xmax=6,
ymin=0,
clip=false,
height=3.3cm,
ymajorgrids=true,
legend cell align=left,
legend columns=3,
legend style={
cells={align=left},
anchor=north,
at={(0.45,1.42)},
draw=none,
fill=none,
column sep=0.5ex,
},
legend image code/.code={%
    \draw[#1,draw] (0cm,-0.04cm) rectangle (0.12cm,0.12cm);
}
]
\addplot+[
    fill = gray, draw=gray, fill opacity=0.5,
    hist={bins=90,data min=2.5, data max=5.5},
    y filter/.expression={y*100/100},
] table[col sep=comma,y=miss] {fig/csv/hit_miss_latency_dist/sd3_h100.csv};
\addplot+[
    fill = green!60!black, draw=green!60!black, fill opacity=0.5,
    hist={bins=90,data min=2.5, data max=5.5},
    y filter/.expression={y*100/100},
] table[col sep=comma,y=10p] {fig/csv/hit_miss_latency_dist/sd3_h100.csv};
\addplot+[
    fill = purple!60!white, draw=purple!60!white, fill opacity=0.5,
    hist={bins=90,data min=2.5, data max=5.5},
    y filter/.expression={y*100/100},
] table[col sep=comma,y=20p] {fig/csv/hit_miss_latency_dist/sd3_h100.csv};
\addplot+[
    fill = orange!60!white, draw=orange!60!white, fill opacity=0.5,
    hist={bins=90,data min=2.5, data max=5.5},
    y filter/.expression={y*100/100},
] table[col sep=comma,y=30p] {fig/csv/hit_miss_latency_dist/sd3_h100.csv};
\addplot+[
    fill = red!60!white, draw=red!60!white, fill opacity=0.5,
    hist={bins=90,data min=2.5, data max=5.5},
    y filter/.expression={y*100/100},
] table[col sep=comma,y=40p] {fig/csv/hit_miss_latency_dist/sd3_h100.csv};
\addplot+[
    fill = blue!60!white, draw=blue!60!white, fill opacity=0.5,
    hist={bins=90,data min=2.5, data max=5.5},
    y filter/.expression={y*100/100},
] table[col sep=comma,y=50p] {fig/csv/hit_miss_latency_dist/sd3_h100.csv};



\end{axis}
\end{tikzpicture}
        \caption{SD3 on H100}
        \label{fig:sd3-h100}
    \end{subfigure}

    \caption{Approximate cache latency distribution ($n=100$). }
    \label{fig:latency-distribution}
\end{figure}
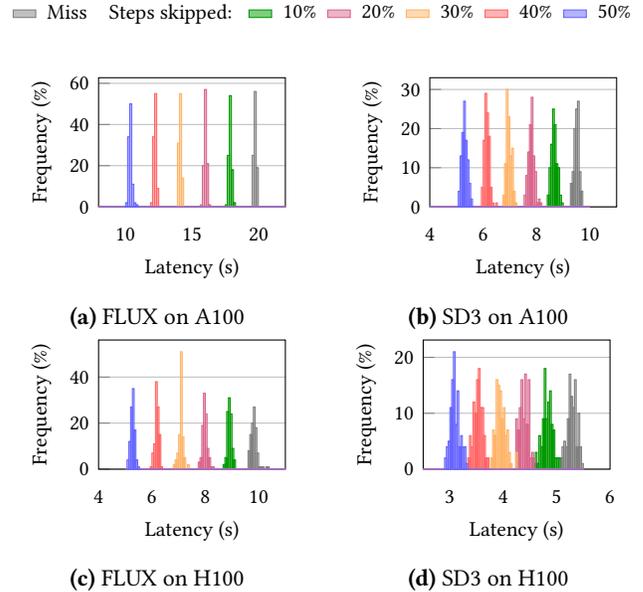

Similar to caches in file systems or processors, approximate caching for diffusion models also aims at reducing execution latency. 
We first analyze its latency to better understand its security implications. 
We set up a text-to-image diffusion serving system on a remote server using the Runpod cloud \cite{runpod}.
The server is over \SI{600}{\kilo\meter} away from our client, ensuring a realistic cloud serving connection. 
We implement NIRVANA~\cite{nirvana}, the state-of-the-art approximate cache from Adobe. 
We follow its configuration that caches every 10\,\% of the denoising steps until 50\,\%. 
We evaluate two popular commercial open-sourced diffusion models, Stable Diffusion 3 Median (SD3) \cite{sd3} and FLUX \cite{labs2025flux1kontextflowmatching,flux2024} with two types of GPU, A100 and H100 (both have \SI{80}{\giga\byte} of memory).
Both diffusion models have 30 denoising steps.

For each diffusion model and GPU type, we randomly select 100 prompts and send requests from our local PC to reflect the impact of network fluctuation. 
Each request is evaluated with different numbers of skipped steps. 
In comparison, the deviation of H100 results is more significant than A100, owing to its faster generation, amplifying the impact of network fluctuation. 
Nonetheless, on both platforms, skipping even just 10\,\% of the steps yields a clear latency reduction, indicating the feasibility of a timing channel.
Moreover, the latency differences between each level of skipped steps are also distinguishable on both platforms. 
As introduced in \Cref{subsec:approx_cache}, the number of skipped steps is correlated with the similarity between the new prompt and the cached prompt. 
In the rest of this paper, we use an H100 for the experiments as it is a more challenging platform for timing channels.

\begin{primitivebox}{box:primitive_1}{}
Because an approximate cache reduces generation latency on cache hits, it enables attackers to determine whether their prompts hit the cache. 
In addition, the attacker can use the magnitude of the latency reduction to infer the similarity between their prompt and the cached prompt they hit. 
\end{primitivebox}

\subsubsection{Similarity of Image Generations}
\label{subsubsec:similarity-generation}
A cache hit reuses an intermediate result from a prior prompt. Therefore, such reuse likely retains information from a previous generation. 
According to previous studies, the diffusion model determines the image's layout at an early stage \cite{nirvana,xia2025modmefficientservingimage, fisedit}, implying that reusing the same cached prompts reduces output diversity.
Although the details may change after multiple denoising steps, the structure of the images remains similar to the cached state.

\begin{figure}[t]
  \centering
  \begin{subfigure}[b]{1\linewidth}
    \centering
    \begin{tikzpicture} 
    \begin{axis}[%
    width=1\hsize,
    style={font=\footnotesize},
    hide axis,
    xmin=10,
    xmax=50,
    ymin=0,
    ymax=0.4,
    legend columns=7,
    column sep=1ex,
    legend image post style={scale=0.5}, 
    legend style={
        cells={align=center},
        anchor=south,
        at={(0.5,0.5)},
        draw=none,
        fill=none,
        column sep=0.5ex,
    },
    ]
    \addlegendimage{area legend, fill = blue, draw=blue, fill opacity=0.4,draw opacity=0.4}
    \addlegendentry{From Same Cache};
    \addlegendimage{area legend, fill = red, draw=red, fill opacity=0.4,draw opacity=0.4}
    \addlegendentry{From Different Cache};

    \end{axis}
\end{tikzpicture}
  \end{subfigure}
  \begin{subfigure}[b]{0.48\linewidth}
      \begin{tikzpicture}
\begin{axis}[
ybar,
style={font=\footnotesize},
xlabel={SSIM score},
ylabel={Frequency (\%)},
width=1\linewidth,
scaled y ticks=false,
xtick pos=bottom,
ytick pos=left,
xtick style={draw=none},
xmin=0,
xmax=1,
ymin=0,
ymax=20,
clip=false,
height=3.3cm,
ymajorgrids=true,
legend cell align=left,
legend columns=3,
legend style={
cells={align=left},
anchor=north,
at={(0.45,1.42)},
draw=none,
fill=none,
column sep=0.5ex,
},
legend image code/.code={%
    \draw[#1,draw] (0cm,-0.04cm) rectangle (0.12cm,0.12cm);
}
]
\addplot+[
    fill = blue, draw=blue, fill opacity=0.4,draw opacity=0.4,
    hist={bins=60,data min=0, data max=1},
    y filter/.expression={y*100/100},
] table[col sep=comma,y=flux same] {fig/csv/ssim_dist.csv};
\addplot+[
    fill = red, draw=red, fill opacity=0.4,draw opacity=0.4,
    hist={bins=60,data min=0, data max=1},
    y filter/.expression={y*100/100},
] table[col sep=comma,y=flux diff] {fig/csv/ssim_dist.csv};



\end{axis}
\end{tikzpicture}
      \caption{FLUX}
  \end{subfigure}
    \hfill
  \begin{subfigure}[b]{0.48\linewidth}
      \begin{tikzpicture}
\begin{axis}[
ybar,
style={font=\footnotesize},
xlabel={SSIM score},
ylabel={Frequency (\%)},
width=1\linewidth,
scaled y ticks=false,
xtick pos=bottom,
ytick pos=left,
xtick style={draw=none},
xmin=0,
xmax=1,
ymin=0,
ymax=15,
clip=false,
height=3.3cm,
ymajorgrids=true,
legend cell align=left,
legend columns=3,
legend style={
cells={align=left},
anchor=north,
at={(0.45,1.42)},
draw=none,
fill=none,
column sep=0.5ex,
},
legend image code/.code={%
    \draw[#1,draw] (0cm,-0.04cm) rectangle (0.12cm,0.12cm);
}
]
\addplot+[
    fill = blue, draw=blue, fill opacity=0.4,draw opacity=0.4,
    hist={bins=60,data min=0, data max=1},
    y filter/.expression={y*100/100},
] table[col sep=comma,y=sd3 same] {fig/csv/ssim_dist.csv};
\addplot+[
    fill = red, draw=red, fill opacity=0.4,draw opacity=0.4,
    hist={bins=60,data min=0, data max=1},
    y filter/.expression={y*100/100},
] table[col sep=comma,y=sd3 diff] {fig/csv/ssim_dist.csv};



\end{axis}
\end{tikzpicture}
      \caption{SD3}
  \end{subfigure}
  \caption{\label{fig:ssim-distribution} SSIM distribution ($n=100$). }
\end{figure}
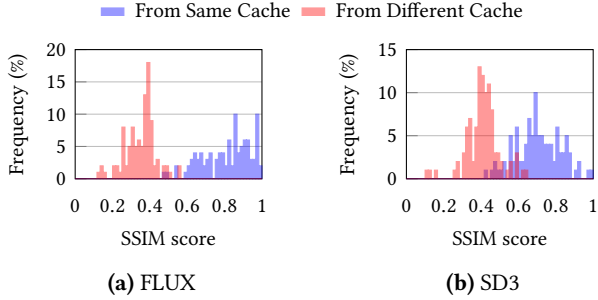

We evaluate such similarity by generating 100 images from different cached states and measuring their Structural Similarity Index Measure (SSIM), which is a common metric to measure the similarity between two images. 
We evaluated both SD3 and FLUX models.
\Cref{fig:ssim-distribution} shows the distribution of SSIM scores between two images.
If two prompts hit the same cached prompt, their SSIM score tends to be higher compared to those that hit different cached prompts.
The difference is more prominent in FLUX, as it is a more powerful model that constructs the layout at very early stages.

\begin{primitivebox}{box:primitive_2}{}
Images generated from the same approximate cache have higher similarity. 
Attackers can use the generated image to infer the original prompt that their prompt hits.
\end{primitivebox}

\section{Remote Covert Channel} \label{sec:covert}
A covert channel is a secret channel for two parties to exchange secret messages. 
In this section, we demonstrate that a sender and a receiver can establish a remote  covert channel through the approximate cache of text-to-image diffusion models to asynchronously transmit messages. 
This is done by sending specially-crafted prompts to leave a message in the cache and retrieving it later. 
Compared to synchronous covert channels \cite{KernelSnitch,sihang23usenixsidechannel,Kurth_SP_2020_NetCAT,DUPEFS,NetSpectre} that require both parties to simultaneously communicate, an asynchronous covert channel through text-to-image generation is highly stealthy.

\subsection{Attack Model} \label{subsec:covert_channel_attack_model}
We assume that the image generation service deploys the approximate caching technique for better efficiency.
When a user sends a prompt, the service caches the prompt and intermediate states of the generation. 
We assume that the sender and the receiver do not have direct communication channels. Instead, they can only access the same remote diffusion-model-based image generation service through prompts. 
The sender and the receiver also do not have any knowledge about the existing cached prompts. 



\begin{figure}
    \centering
    \includegraphics[width=1\linewidth]{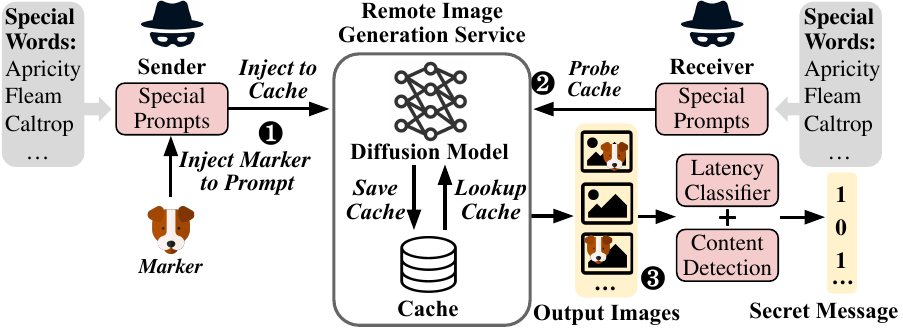}
    \caption{The setup of the remote covert channel, which is based on approximate cache in diffusion models.}
    \label{fig:covert_channel_scenario}
\end{figure}

\begin{figure*}
  \begin{center}
  \includegraphics[width=\linewidth]{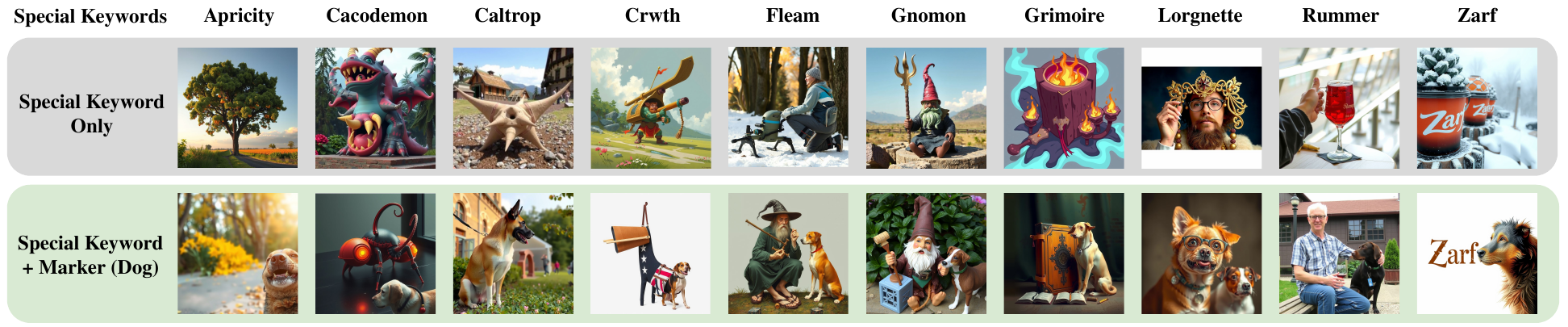}
  \end{center}
  \vspace{-1em}
  \caption{\label{fig:convert_channel_example} Covert channel examples (image generated by FLUX).
  }
\end{figure*}

\subsection{Covert Channel Design}
\label{subsec:convert-channel-design}

Establishing an asynchronous covert channel, through which the sender leaves a message and the receiver retrieves it at a later time, requires accurate identification of whether the receiver’s prompt hits a cached prompt left by the sender.
\Cref{box:primitive_1} indicates that a prompt that hits a previous prompt can be identified through its image generation latency.
As an approximate cache contains cached prompts from various users, it is difficult to determine whether the cached prompt that the receiver's prompt hit was left by the sender.
Therefore, we use a special keyword (\eg uncommon words) to differentiate the sender's prompt, which the receiver can use to retrieve the cached prompt.
However, the approximate cache may treat a prompt as a hit if another prompt is similar to other parts of the receiver's prompt, even if the sender did not inject such a prompt into the approximate cache. 
\Cref{box:primitive_2} indicates that the output image can be used to infer whether the new prompt hits the original cached prompt, as they have higher similarity than other generated images. 
Based on both attack primitives, we additionally include a special marker.  
When leaving the message, the sender combines the marker with the special keyword in the prompt.
Then, the receiver can probe the cache using the keyword and perform a two-level detection to confirm whether the prompt hits a cached prompt (using latency) and whether the hit is on the sender's prompt (by detecting the marker in the output image). 

\Cref{fig:covert_channel_scenario} demonstrates the covert channel process.
First, the sender injects prompts to the approximate cache with special keywords (\scalebox{0.8}{\circled{1}}).
Then, the receiver sends prompts with the same keywords (\scalebox{0.8}{\circled{2}}).
Finally, the receiver uses the two-level detection to determine whether they hit the sender's prompt (\scalebox{0.8}{\circled{3}}). A hit or a miss transmits a bit of 1 or 0, respectively. 
Specifically, the covert channel has the following components.

\noindent\textbf{Unique prompt generator.}
We first use an LLM to generate a set of special, ancient words that are no longer used today, without any prior knowledge about the prompts in the test dataset. 
These special words are preknowledge shared by the sender and the receiver (e.g., shared offline before transmission). 
Then, both the sender and receiver use an LLM to construct prompts.
On the sender side, the Llama model constructs a prompt for each special word with the marker.
On the receiver side, the Llama model constructs a prompt for each special word to probe the cache. 

\noindent\textbf{Latency classifier.}
The receiver profiles the generation latency of the serving system and uses the average latency of hit and miss cases to build a classifier that determines whether their probing prompt hits any cached prompts.

\noindent\textbf{Content detector.}
The receiver uses a content detection model to recognize the predefined marker in the output image.
If recognized, the receiver's prompt hits the sender's prompt.

\subsection{Setup} \label{subsec:covert_channel_setup}

\noindent\textbf{Platform.} We use the approximate cache system and cloud platform introduced in \Cref{subsubsec:cache_timing}.

\noindent\textbf{Model and Dataset.}
We select the state-of-the-art text-to-image diffusion model, FLUX \cite{flux2024}, as the primary model for evaluation.
We additionally evaluate a smaller model, Stable Diffusion 3 Medium (SD3) \cite{sd3}. 
We use \diffusiondb{} as the main dataset which contains real-world user prompts and use \lexica{} as a secondary dataset which contains a smaller number of prompts but in different styles. 
While we only choose prompts in English, in practice, other languages can also be used for this covert channel. 
We configure \SI{100}{\giga\byte} cache size for \diffusiondb{} and \SI{1}{\giga\byte} for \lexica{} due to their number of prompts and deploy the same cache replacement policy as NIRVANA \cite{nirvana} as default. 

\noindent\textbf{Correctness Metric.}
The success rate depends on two factors:
\begin{itemize}[leftmargin=*,noitemsep,partopsep=0pt,topsep=0pt,parsep=0pt]
  \item \textbf{Insertion success rate} measures the rate of successful cache insertion by the sender. If the sender posts a request but hits the cache, the cache system will not insert a new cache, leading to incorrect message transmission.
  \item \textbf{Detection success rate} measures the receiver's rate of successful retrieval of prompts with special words and the associated detection of the marker, once the sender has injected the special prompts into the cache successfully.
\end{itemize} 


\noindent\textbf{Sender and Receiver Setup. }
We use a Llama-3 70B model as the LLM to construct 10 special keywords and the corresponding unique prompts.
\Cref{fig:convert_channel_example} illustrates these keywords and images generated by FLUX for message transmission. 
These extremely uncommon words increase the chance that the receiver’s prompts hit the sender’s cached prompts. 
We use ``dog'' as the marker for FLUX (this example) and ``McDonald's'' for SD3.
This is because FLUX is a more powerful model that can render complicated objects from the cached states compared to SD3.
If the receiver does not hit the sender's prompts, the output images resemble those in the first row; otherwise, the receiver obtains outputs like those in the second row. 
DINOv2 \cite{oquab2023dinov2} is used to detect the marker in the receiver's generated images, which is the same method of object detection in Silent Branding \cite{jang2025silentbranding}. 

\noindent\textbf{Baseline. }
We introduce a baseline that only uses generation latency of the prompt with special words to determine cache hit or miss, without using any marker.

\subsection{Results}
 
This section presents the evaluation results.

\subsubsection{Overall covert channel accuracy}
We evaluate covert channel 100 times on both FLUX and SD3.
Both sender and receiver send requests to the remote cloud. We assume that the cache system has all the unique prompts from \diffusiondb{}-2M. 
\Cref{tab:success_rate_breakdown} shows that the covert channel achieves over 97.8\,\% and 95.8\,\% accuracy in FLUX and SD3, respectively. FLUX achieves a higher and more stable accuracy than SD3.
This is because FLUX is a more powerful model, capable of generating clearer objects even in early denoising steps. Thus, the cached intermediate states in FLUX are more likely to preserve these objects. 

\begin{table}
    \caption{Accuracy breakdown ($n=100$).}
    \label{tab:success_rate_breakdown}
    \centering
    \small
    \setlength{\tabcolsep}{2.5pt}
    \begin{tabular}{c C{4cm} C{3cm}}
    \toprule
        Model  &  \makecell{Overall Accuracy\\(Latency + Content Classifier)}  & Latency-only Classifier \\
    \midrule
        FLUX & 97.8\,\%, $\sigma$=2.5\,\% & 94.6\,\%, $\sigma$=13.0\,\% \\
        SD3  & 95.8\,\%, $\sigma$=3.6\,\% & 90.8\,\%, $\sigma$=16.1\,\% \\
        \bottomrule
    \end{tabular}
\end{table}

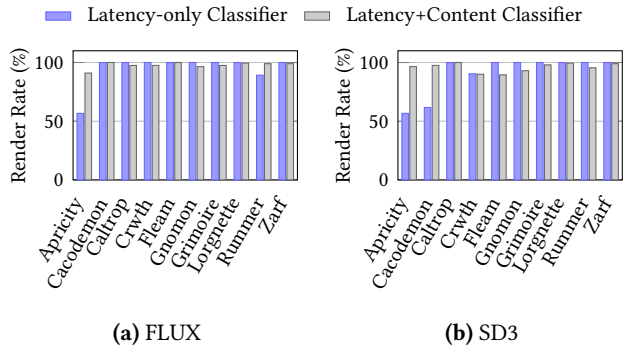
\begin{figure}
    \centering
    \begin{subfigure}[b]{1\linewidth}
        \centering
        \begin{tikzpicture} 
    \begin{axis}[%
    width=1\hsize,
    style={font=\footnotesize},
    hide axis,
    xmin=10,
    xmax=50,
    ymin=0,
    ymax=0.4,
    legend columns=3,
    column sep=1ex,
    legend image post style={scale=0.55}, 
    legend style={
        cells={align=center},
        anchor=south,
        at={(0.5,0.5)},
        draw=none,
        fill=none,
        column sep=1ex,
    },
    ]
    \addlegendimage{area legend, draw=blue!60!white,fill=blue!40!white}
    \addlegendentry{Latency-only Classifier};
    \addlegendimage{area legend, draw=black!60!white,fill=black!20!white}
    \addlegendentry{Latency+Content Classifier};
    \end{axis}
\end{tikzpicture}
    \end{subfigure}
    \begin{subfigure}[b]{0.49\linewidth}
        \begin{tikzpicture}
\begin{axis}[
    style={font=\footnotesize},
    ybar,
    ylabel shift={-5pt},
    bar width=2.5pt,
    clip=false,
    height=3.3cm,
    width=1.1\hsize,
    xmin=-0.5,
    xmax=9.5,
    ymin=0,
    ymax=110,
    xtick = {0,1,2,3,4,5,6,7,8,9},
    xticklabels={Apricity, Cacodemon, Caltrop, Crwth, Fleam, Gnomon, Grimoire, Lorgnette, Rummer, Zarf},
    xticklabel style={rotate=60,anchor=east},
    ylabel={Render Rate (\%)},
    xtick style={draw=none},
    ymajorgrids=true,
    legend columns=1,
    legend cell align=left,
    legend style={
        cells={align=left},
        anchor=south,
        at={(1.2,0.5)},
        draw=none,
        fill=none,
        column sep=0.5ex,
    },
    legend image code/.code={%
    \draw[#1, draw] (0cm,-0.05cm)
    rectangle (0.2cm,0.1cm);
    }
]

\addplot [draw=blue!60!white,fill=blue!40!white,bar shift=-1.5pt] table[col sep=comma, meta=keyword, x expr=\coordindex, y=flux_latency-only] {fig/csv/attack1/keyword_fp.csv};
\addplot [draw=black!60!white,fill=black!20!white,bar shift=1.5pt] table[col sep=comma, meta=keyword, x expr=\coordindex, y=flux_content+latency] {fig/csv/attack1/keyword_fp.csv};

\end{axis}

\end{tikzpicture}
        \caption{FLUX}
    \end{subfigure}
    \hfill
    \begin{subfigure}[b]{0.49\linewidth}
        \begin{tikzpicture}
\begin{axis}[
    style={font=\footnotesize},
    ybar,
    ylabel shift={-5pt},
    bar width=2.5pt,
    clip=false,
    height=3.3cm,
    width=1.1\hsize,
    xmin=-0.5,
    xmax=9.5,
    ymin=0,
    ymax=110,
    xtick = {0,1,2,3,4,5,6,7,8,9},
    xticklabels={Apricity, Cacodemon, Caltrop, Crwth, Fleam, Gnomon, Grimoire, Lorgnette, Rummer, Zarf},
    xticklabel style={rotate=60,anchor=east},
    ylabel={Render Rate (\%)},
    xtick style={draw=none},
    ymajorgrids=true,
    legend columns=1,
    legend cell align=left,
    legend style={
        cells={align=left},
        anchor=south,
        at={(1.2,0.5)},
        draw=none,
        fill=none,
        column sep=0.5ex,
    },
    legend image code/.code={%
    \draw[#1, draw] (0cm,-0.05cm)
    rectangle (0.2cm,0.1cm);
    }
]

\addplot [draw=blue!60!white,fill=blue!40!white,bar shift=-1.5pt] table[col sep=comma, meta=keyword, x expr=\coordindex, y=sd3_latency-only] {fig/csv/attack1/keyword_fp.csv};
\addplot [draw=black!60!white,fill=black!20!white,bar shift=1.5pt] table[col sep=comma, meta=keyword, x expr=\coordindex, y=sd3_content+latency] {fig/csv/attack1/keyword_fp.csv};

\end{axis}

\end{tikzpicture}
        \caption{SD3}
    \end{subfigure}
    \caption{Render success rate of each keyword ($n=100$).}
    \label{fig:keyword_accuracy}
\end{figure}

We further explore the accuracy breakdown.
Incorrect transmission can potentially come from two sources.
First, the sender's prompt may hit an existing cached prompt, without being cached in the serving system. Consequently, the receiver always receives a bit of 0.
Our experiment shows a 100\,\% cache insertion success rate because the sender's prompts are not similar to either \lexica{} or \diffusiondb{}-2M. 
Second, the \textit{latency classifier} and \textit{content detector} in two-level detection process can also introduce inaccuracies.
\Cref{tab:success_rate_breakdown} shows the \textit{latency classifier}'s average accuracy, which is high (94.6\% and 90.8\% for FLUX and SD3, respectively) but varies by keyword.
Misclassifications occur most often when the sender omits a prompt for a keyword (to transmit a 0), where the receiver may hit other cached prompts and cause false positives.
\Cref{fig:keyword_accuracy} presents the \textit{latency classifier}'s accuracy for all 10 keywords.
It achieves less than 60\,\% accuracy for some keywords (\eg{} \textit{Apricity}), likely because prompts with these keywords have high similarity scores with normal prompts. 
In contrast, including the \textit{content classifier} ensures a more consistent accuracy in both models.
In a few cases, the \textit{content classifier} slightly lowers the accuracy because the model fails to render the marker clearly. This is more prominent in the weaker model, SD3 (\eg{} \textit{Fleam}).

\subsubsection{Lifetime of Sender's Cache}
\label{subsubsec:covert_lifetime}
We evaluate the lifetime of the sender’s cache entries, \ie{} how long an injected prompt remains cached before eviction.
We follow the same setup as NIRVANA \cite{nirvana}, where a prompt is injected only if it does not hit any cached prompt. 
We replay the trace in \diffusiondb{}-Large with 14M prompts according to the timestamps and use FLUX as the image generation model.
We insert the sender's prompts once the approximate cache is full, and count the time and number of requests the service received until eviction.

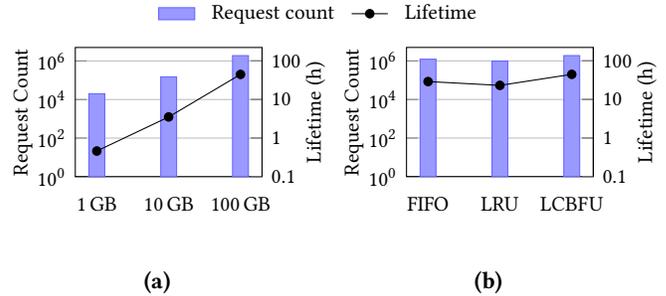
\begin{figure}
\centering
\begin{subfigure}[b]{1\linewidth}
    \centering
    \begin{tikzpicture} 
    \begin{axis}[%
    style={font=\footnotesize},
    hide axis,
    xmin=10,
    xmax=50,
    ymin=0,
    ymax=0.4,
    legend columns=3,
    column sep=1ex,
    legend image post style={scale=0.8}, 
    legend style={draw=none,nodes={anchor=east}},
    ]
    \addlegendimage{area legend, draw=blue!60!white,fill=blue!40!white}
    \addlegendentry{Request count};
    \addlegendimage{draw=black, mark=*, mark options={black, scale=1}}
    \addlegendentry{Lifetime};
    \end{axis}
\end{tikzpicture}
\end{subfigure}
\begin{subfigure}[b]{0.48\linewidth}
    \begin{tikzpicture}
\begin{axis}[
    style={font=\footnotesize},
    ybar,
    ymode=log,
    ylabel shift={-5pt},
    bar width=6pt,
    clip=false,
    height=3.3cm,
    width=1\hsize,
    xmin=-0.3,
    xmax=2.3,
    ymin=1,
    ymax=5e+06,
    xtick = {0,1,2},
    xticklabels={\SI{1}{\giga\byte},\SI{10}{\giga\byte},\SI{100}{\giga\byte} },
    ylabel={Request Count},
    xtick style={draw=none},
    ymajorgrids=true,
    legend columns=1,
    legend cell align=left,
    legend style={
        cells={align=left},
        anchor=south,
        at={(1.2,0.5)},
        draw=none,
        fill=none,
        column sep=0.5ex,
    },
    legend image code/.code={%
    \draw[#1, draw] (0cm,-0.05cm)
    rectangle (0.2cm,0.1cm);
    }
]

\addplot [draw=blue!60!white,fill=blue!40!white] table[col sep=comma, meta=size, x expr=\coordindex, y=req count] {fig/csv/attack1/cache_size.csv};

\end{axis}

\begin{axis}[
    style={font=\footnotesize},
    ymode=log,
    ylabel shift={-5pt},
    clip=false,
    height=3.3cm,
    width=1\hsize,
    xmin=-0.3,
    xmax=2.3,
    ymin=0.1,
    ymax=220,
    xtick = {0,1,2},
    axis x line=none,
    axis y line*=right,
    ytick = {0.1,1,10,100},
    yticklabels={0.1,1,10,100},
    ylabel={Lifetime (h)},
    yminorticks=false,
    legend columns=1,
    legend cell align=left,
    legend style={
        cells={align=left},
        anchor=south,
        at={(1.2,0.5)},
        draw=none,
        fill=none,
        column sep=0.5ex,
    },
    legend image code/.code={%
    \draw[#1, draw] (0cm,-0.05cm)
    rectangle (0.2cm,0.1cm);
    }
]

\addplot+ [black, mark=*, mark options={black, scale=.8},] table[col sep=comma, meta=size, x expr=\coordindex, y=time] {fig/csv/attack1/cache_size.csv};

\end{axis}

\end{tikzpicture}
    \caption{}\label{tab:conver-channel-lifespan-size}
\end{subfigure}
\hfill
\begin{subfigure}[b]{0.48\linewidth}
    \begin{tikzpicture}
\begin{axis}[
    style={font=\footnotesize},
    ybar,
    ymode=log,
    ylabel shift={-5pt},
    bar width=6pt,
    clip=false,
    height=3.3cm,
    width=1\hsize,
    xmin=-0.3,
    xmax=2.3,
    ymin=1,
    ymax=5e+06,
    xtick = {0,1,2},
    xticklabels={FIFO, LRU, LCBFU},
    ylabel={Request Count},
    xtick style={draw=none},
    ymajorgrids=true,
    legend columns=1,
    legend cell align=left,
    legend style={
        cells={align=left},
        anchor=south,
        at={(1.2,0.5)},
        draw=none,
        fill=none,
        column sep=0.5ex,
    },
    legend image code/.code={%
    \draw[#1, draw] (0cm,-0.05cm)
    rectangle (0.2cm,0.1cm);
    }
]

\addplot [draw=blue!60!white,fill=blue!40!white] table[col sep=comma, meta=policy, x expr=\coordindex, y=req count] {fig/csv/attack1/replacement_policy.csv};

\end{axis}

\begin{axis}[
    style={font=\footnotesize},
    ymode=log,
    ylabel shift={-5pt},
    clip=false,
    height=3.3cm,
    width=1\hsize,
    xmin=-0.3,
    xmax=2.3,
    ymin=0.1,
    ymax=220,
    xtick = {0,1,2},
    axis x line=none,
    axis y line*=right,
    ytick = {0.1,1,10,100},
    yticklabels={0.1,1,10,100},
    ylabel={Lifetime (h)},
    yminorticks=false,
    legend columns=1,
    legend cell align=left,
    legend style={
        cells={align=left},
        anchor=south,
        at={(1.2,0.5)},
        draw=none,
        fill=none,
        column sep=0.5ex,
    },
    legend image code/.code={%
    \draw[#1, draw] (0cm,-0.05cm)
    rectangle (0.2cm,0.1cm);
    }
]

\addplot+ [black, mark=*, mark options={black, scale=0.8},] table[col sep=comma, meta=policy, x expr=\coordindex, y=time] {fig/csv/attack1/replacement_policy.csv};

\end{axis}

\end{tikzpicture}
    \caption{}\label{tab:conver-channel-lifespan-policy}
\end{subfigure}
\caption{Sender's cache lifetime under (a) different cache sizes (LCBFU) and (b) replacement policies (\SI{100}{\giga\byte} cache).}
\end{figure}

\Cref{tab:conver-channel-lifespan-size} shows the lifetime under different cache sizes.
As the cache size increases, the lifetime of injected cached prompts increases significantly. 
\Cref{tab:conver-channel-lifespan-policy} shows the lifetime under three replacement policies: 
LCBFU and FIFO are the default replacement policies in prior approximate caching systems \cite{nirvana, xia2025modmefficientservingimage}; LRU is a popular replacement policy.
We observe that LCBFU retains the sender's prompts in the cache for the longest time because it manages caches in a finer granularity, effectively allowing for more cache entries.
LRU evicts the sender's prompt earliest, as these keywords barely receive hits from normal users.
Overall, the sender's prompts last in the cache for a long time (up to 44 hours), making the channel stealthy.

\subsubsection{Time and Resource Consumptions}
\label{subsubsec:covert_resource}

We further conduct an experiment to measure the resource consumption of the covert channel. 
\Cref{tab:time-consumption} shows the results when running the models on an H100. 
Because every image represents 1 bit, sending and receiving 1 bit both take 1 prompt. 
In our setup, each image has \SIx{16384} tokens due to the 8$\times$ downsample ratio from the 1024 $\times$ 1024 output resolution in both diffusion models. 
Therefore, sending and receiving 1 bit both consume \SIx{16384} tokens. 
We also evaluate the time for sending and receiving 1 bit.
Sending 1 bit takes $1.20\times$ and $1.23\times$ more time compared to receiving one bit in FLUX and SD3, respectively. 
This difference is because sending incurs cache misses, while receiving benefits from cache hits.
Although the bandwidth is lower than synchronous covert channels \cite{sihang23usenixsidechannel,Kurth_SP_2020_NetCAT}, it is comparable to other asynchronous covert channels \cite{Martin_NDSS_2022_Remote}.

\begin{table}
    \caption{Time and resource consumptions per bit on H100.}
    \label{tab:time-consumption}
    \centering
    \small
    \setlength{\tabcolsep}{4pt}
    \begin{tabular}{c c c}
    \toprule
        Model & FLUX & SD3 \\
    \midrule
    \# Prompts (Sending/Receiving)  & 1 & 1 \\
    \# Tokens (Sending/Receiving)  & \SIx{16384} & \SIx{16384} \\
    Sending Time (s) & 9.8 & 5.3 \\
    Receiving Time (s) & 8.2 & 4.3 \\
    \bottomrule
    \end{tabular}
\end{table}

\subsection{Discussions} \label{subsec:covert_channel_discussion}

Compared to prior attacks on diffusion models that transmits secret messages \cite{chen2025parasitesteganographybasedbackdoorattack,mahfuz2025psyducktrainingfreesteganographylatent,Pulsar,CRoSS,LDStega}, this covert channel does not modify or train the diffusion model. 
Moreover, instead of using the timing channel alone, like remote covert channels on conventional caches \cite{KernelSnitch,sihang23usenixsidechannel,Kurth_SP_2020_NetCAT,DUPEFS,NetSpectre}, this covert channel analyzes the generation output to achieve a higher success rate. 
As discussed in \Cref{subsubsec:covert_resource}, this covert channel requires a total of 2 prompts when sending and receiving 1 bit, leading to a lower bandwidth than some synchronous covert channels \cite{sihang23usenixsidechannel,Kurth_SP_2020_NetCAT} and higher resource consumptions. 
However, its asynchronous property allows the sender's message to remain in the approximate cache for days, enabling the receiver to obtain the message at a later time. 
This enhances the stealthiness of this channel and makes it harder to be detected.
Moreover, in case of unstable networking and delayed response, the \textit{content classifier} is still effective as it can still check the marker in the output image without relying on latency measurements. 

\section{\stealing{}: Prompt Stealing Attack}
\label{sec:stealing}
As prior work has shown, generating high-quality images requires prompt engineering and careful design of modifiers~\cite{xinyue_prompt_stealing,cross_modal_prompt_inversion}.
In this section, we introduce \stealing{}, a prompt stealing attack, where the attacker's goal is to steal other users' prompts (e.g., for commercial usage) from the approximate cache to reproduce their image generation. 

\subsection{Attack Model}
\label{subsec:stealing_attack_mode}

We assume the same image generation serving system as \Cref{subsec:covert_channel_attack_model}. 
Similarly, we assume the attacker has no additional privileges than other normal users, \ie zero knowledge of other users' prompts or generated images and they can only send requests to those generation services to get images. 
The attacker is also unaware of the cache size, but only identifies which probing prompts hit the same cached prompt and recovers the corresponding prompt.


\begin{figure}[t]
  \begin{center}
  \includegraphics[width=\linewidth]{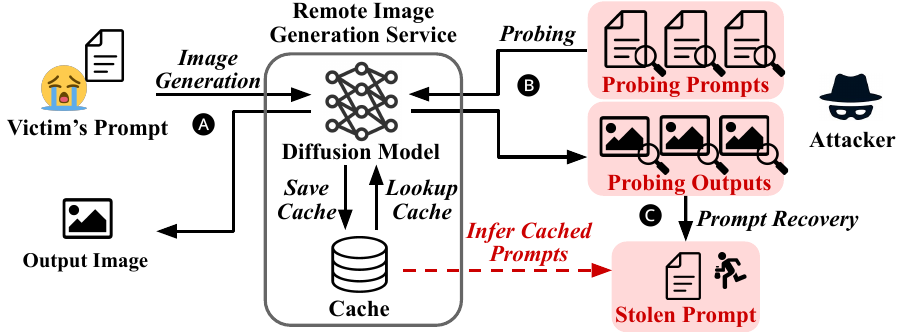}
  \end{center}
  \caption{\label{fig:prompt_steal_scenario} The setup and overview of \stealing{}.}
\end{figure}

\begin{figure*}
  \begin{center}
  \includegraphics[width=\linewidth]{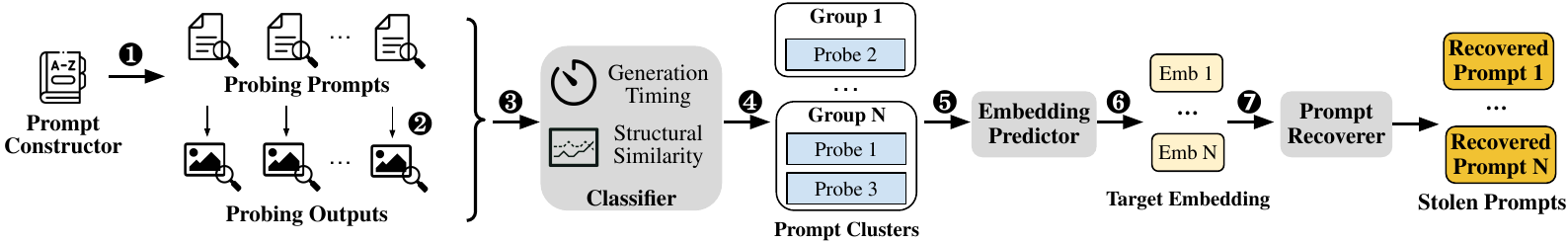}
  \end{center}
  \caption{\label{fig:prompt_steal_overview} Design details of \stealing{}. }
\end{figure*}

\subsection{Attack Design}
\label{subsec:stealing_design}

\Cref{fig:prompt_steal_scenario} illustrates an overview \stealing{} attack.
The victim sends a prompt to the image generation service and the prompt is cached (\scalebox{0.7}{\circled{A}}).  
The attacker later sends probing prompts to the service (\scalebox{0.7}{\circled{B}}) and recovers the victim's prompt by analyzing the probing prompts (\scalebox{0.7}{\circled{C}}).

\subsubsection{Challenges}
The attacker can distinguish a probing prompt's hit/miss through latency (\Cref{box:primitive_1}). 
However, the similarity between the probing prompt and the target prompt can be low (\eg{} only 0.65).
Thus, a single probing prompt is insufficient for recovery.
Using multiple probing prompts should improve the accuracy but two challenges need to be solved. 
First, it is difficult to determine which  cached prompt a probing prompt hits.
Approximate caching retrieves prompts by similarity, which introduces uncertainties. 
A misclassification on the target prompt degrades the recovery accuracy. 
Second, the embedding space does not directly map to words.
Therefore, even with an accurate classification of hits, converting multiple probing prompts that are similar in the embedding space back to the original prompt remains challenging. 

\subsubsection{Solution}
To address these challenges, \stealing{} follows the approach in \Cref{fig:prompt_steal_overview}. 
First, we design a \textit{prompt constructor} to construct probing prompts (\scalebox{0.8}{\circled{1}}) and get corresponding images (\scalebox{0.8}{\circled{2}}). Once the attacker has collected enough probing data, a two-level \textit{classifier} (\scalebox{0.8}{\circled{3}}) clusters these prompt-image pairs into multiple groups, where probing prompts in the same group hit the same target prompt (\scalebox{0.8}{\circled{4}}). 
After that, the attacker uses an \textit{embedding predictor} (\scalebox{0.8}{\circled{5}}) to guess the cache embedding that every group hits (\scalebox{0.8}{\circled{6}}). 
The created embedding maintains high similarity to every prompt in the group. 
We consider this embedding as the approximation of the cached prompt's embedding and recover it back to the prompt with a \textit{prompt recoverer} (\scalebox{0.8}{\circled{7}}). 
The recovered prompt is the stolen prompt.
The attacker sends these prompts back to the same generation application to obtain the images. 
We next discuss these components in detail.

\noindent\textbf{Prompt Constructor: }
The \emph{prompt constructor} generates a batch of prompts by sampling a variable number of modifiers from a predefined modifier collection file. 
We use the same modifier pool as a prior study \cite{xinyue_prompt_stealing}.
For each attack trial, the \textit{prompt constructor} starts from a subject and randomly samples modifiers from the pool. 
The number of modifiers per probing prompt varies. After constructing a prompt, the attacker sends it to the image generation service and exploits the latency to determine whether this prompt is a cache hit. 
Once a hit is detected, the attacker collects the prompts for the following processing.
The \textit{prompt constructor} also leverages a filter to prevent the constructed prompts from being too similar (keeping semantic similarity below 0.6 in our experiment) to the hit prompts.
By increasing the diversity of hit prompts, this approach prevents overfitting when training the \textit{embedding predictor}.

\noindent\textbf{Classifier.} 
\stealing{}'s classifier consists of two components: a \textit{timing classifier} that determines if a prompt is a cache hit based on its generation latency and a \textit{structural classifier} that further determines which probing prompts hit the same cached prompt. 
The attacker sends the probing prompts generated by the \emph{prompt constructor} to the image generation service and measures the response latency.
The \textit{timing classifier} places prompts with response time within the cache hit latency threshold (\SI{9.5}{\second} for FLUX and \SI{5.06}{\second} for SD3 on an H100) in a \textit{hit set}. 
According to \Cref{box:primitive_2}, generated images tend to be more similar if they hit the same cached prompts. 
Therefore, to further increase the classifier's accuracy, the \textit{structural classifier} examines the prompts' CLIP similarity to avoid false positives. 
It employs a 3-layer MLP, where the input is the CLIP similarity of two prompts and the SSIM scores of the corresponding images, and the output is whether two prompts hit the same cached prompts.  

\noindent\textbf{Embedding Predictor.}
Once the classifier has grouped all probing prompts, the attacker can then recover each group back to the original cached prompt that they hit.  
The \emph{embedding predictor} is an embedding training process with the Stochastic Gradient Descent (SGD) algorithm. It first creates a random embedding $E$ and computes the similarity scores as a vector $S_{E}$ with each prompt from the group. 
We denote the similarity scores between the cached prompt and the prompts in the group as a vector $S_C$.
Since the attacker can only infer the number of skipped steps based on the generation latency (\Cref{box:primitive_1}) instead of an accurate similarity score, we assume that the number of skipped steps represents the worst-case similarity.  
The \emph{embedding predictor} takes the Mean Squared Error (MSE) between $S_E$ and $S_C$ as the loss function, and iteratively updates the embedding $E$ based on SGD to make it similar to the original cached prompt.

\noindent\textbf{Prompt Recoverer.}
The next step is to convert the embedding to prompts. 
Inspired by prior studies \cite{mokady2021clipcapclipprefiximage}, we design a generative model named \textit{prompt recoverer} to convert embeddings back to prompts. It first maps an embedding from the CLIP embedding space to the GPT-2 space with an MLP-based model.
Then, it uses a 124M-parameter GPT-2 \cite{Radford2019LanguageMA} to reconstruct the prompt word by word.
When training, we do not freeze the parameters of GPT-2 to enhance the alignment between the output and normal users' prompt style.

\begin{figure}[t]
  \begin{center}
  \includegraphics[width=\linewidth]{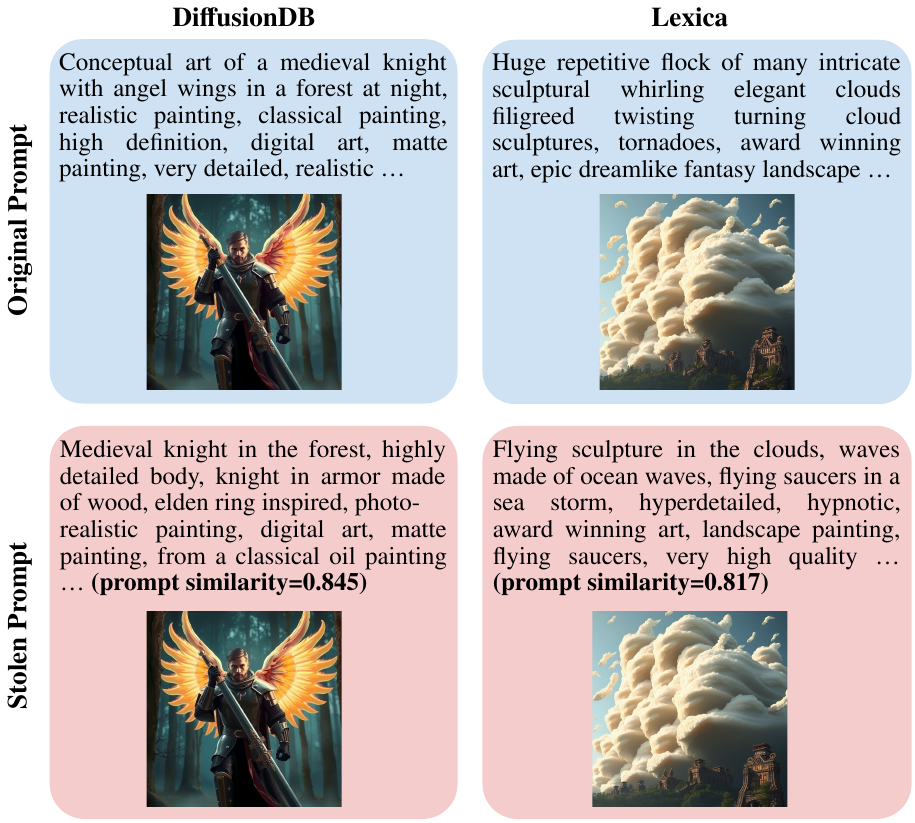}
  \end{center}
  \caption{\label{fig:prompt_steal_example} Examples of \stealing{}.}
\end{figure}

\subsection{Attack Setup}
\label{subsec:stealing_setup}

\noindent\textbf{Platform.} Like before, we use the approximate cache system in \Cref{subsubsec:cache_timing}.

\noindent\textbf{Model and Dataset.}
We follow the same diffusion models and datasets as the covert channel, as described in \Cref{subsec:covert_channel_setup}.
The attacker is unaffected by replacement, as they only probe the approximate cache. 
Therefore, we evaluate this attack using a fixed number of cache entries. 
Due to differences in dataset size, the approximate cache contains 100k entries for \diffusiondb{} and 10k entries for \lexica{}.

\noindent\textbf{\stealing{} Training.} \stealing{} consists of four components as described in \Cref{subsec:stealing_design}.
The \textit{prompt recoverer} and  \textit{classifier} need to be trained. 
We randomly select 20\,\% prompts in \diffusiondb{} to train the two models, and evaluate \stealing{} on the other 80\,\% prompts and the \emph{unseen} dataset \lexica{}. 
Specifically, we select 200 unique prompts in the 20\,\% training set as cached prompts, and use other prompts that are able to hit these cached prompts to generate images.
Using these prompt-image pairs, we train the \textit{classifier} based on their structural and CLIP embedding similarities. 
\textit{Prompt recoverer} takes the CLIP embedding as the input and the original text as the target output.

\noindent\textbf{\stealing{} Probing.} The probing process takes two stages: the first stage constructs the prompts randomly and exploits the latency classifier to collect the hit prompts. Once the hits exceed a certain number (35 in our evaluation), probing moves to the second stage, where embeddings whose similarity to the hit prompts falls within an interval (0.5--0.6 in our evaluation) are sampled. This increases the likelihood of hitting the cached prompt while maintaining prompt diversity.

\noindent\textbf{Baseline.} As our attack model assumes that the attacker has no access to the victim's generated images, previous prompt stealing attacks \cite{xinyue_prompt_stealing} on diffusion models do not work.
Therefore, we use a naive side-channel attack method as the baseline, that only uses the \textit{prompt constructor} to generate similar prompts and a simple \textit{timing classifier} to determine hit/miss, which has been used in prior prompt stealing attacks on LLMs~\cite{zheng2024inputsnatchstealinginputllm,song_TIFC_2025_early_bird}.
We refer to this method as \emph{Naive Probing}.

\noindent\textbf{Accuracy Metrics. }
We evaluate the similarity of stolen prompts and their generated images with the victim's prompts and images using the metrics below:

\begin{itemize}[leftmargin=*,noitemsep,partopsep=0pt,topsep=0pt,parsep=0pt]

\item \textbf{Prompt semantic similarity.} 
The stolen prompt should describe similar content as the target prompt. 
We first evaluate the semantic similarity using cosine similarity in the CLIP embedding space \cite{xinyue_prompt_stealing,cross_modal_prompt_inversion}. This metric is also what \textit{Prompt Recoverer} is optimized for.

\item \textbf{Prompt word-level similarity.}
In addition to the semantic similarity, we also evaluate their word-level similarities using the metrics below:
    \begin{itemize}[leftmargin=*,noitemsep,partopsep=0pt,topsep=0pt,parsep=0pt]
    \item Bilingual Evaluation Understudy (\textbf{BLEU}) \cite{Tan_SP_2025_On,cross_modal_prompt_inversion} is a metric that considers both the word count and the n-gram word sequence count, but does not consider the semantic similarity or synonyms. 
    Specifically, we use the BLEU-1 score to reflect the word-level success rate.
    
    \item Named Entity Recognition Recall (\textbf{NERR}) \cite{wang-wang-2022-sentence} measures the success rate at the entity level. It counts the correct entity numbers in the candidate text compared to the reference text and calculates the recall rate.
    
    \item \textbf{BERTScore} \cite{Tan_SP_2025_On} captures the meaning of words by employing a pretrained BERT model to align words between candidate and reference texts, thus considering synonyms. We use BERTScore F1 as our metric.
    \end{itemize} 

\item \textbf{Image similarity.} The image generated by the stolen prompt should also be similar to the original generation. We evaluate image similarity using the metrics below:
    \begin{itemize}[leftmargin=*,noitemsep,partopsep=0pt,topsep=0pt,parsep=0pt]
    \item Contrastive Language-Image Pre-training (\textbf{CLIP})~\cite{xinyue_prompt_stealing,cross_modal_prompt_inversion} encodes images to embeddings. The cosine similarity between two embeddings reflects the semantic similarity between two images.
    \item Peak Signal-to-Noise Ratio (\textbf{PSNR}) \cite{jang2025silentbranding,CRoSS,sd3} is a pixel-wise image similarity metric. We employ PSNR to quantify the differences between 
    two images.
    \item Structural Similarity Index Measure (\textbf{SSIM}) is the metric used in \Cref{subsubsec:similarity-generation}, which captures the structural similarities between images.
    \end{itemize}

\end{itemize}

\subsection{Results}

\Cref{fig:prompt_steal_example} demonstrates two examples of \stealing{} from \diffusiondb{} and \lexica{}.
The first row shows user prompts and the generated images.
The second row shows \stealing{}'s stolen prompts and the reproduced images.
The stolen prompts are also similar to the user's prompts. Even though there exist word-level differences, from a model's perspective, the two prompts are highly similar. Thus, the generated images are almost identical to the original.

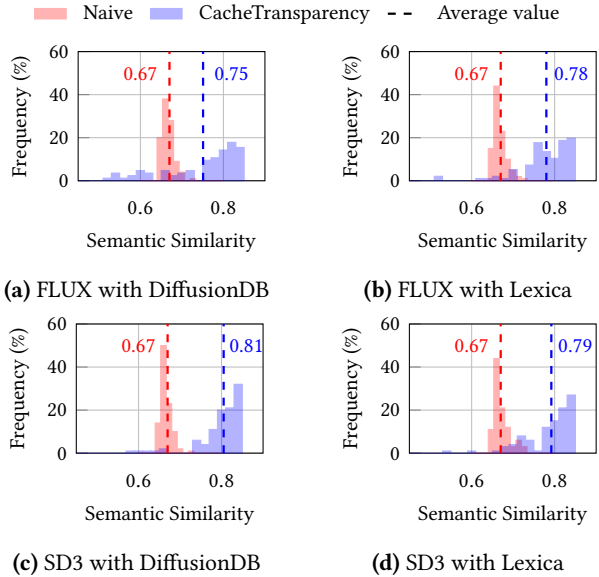
\begin{figure}[t]
    \centering
    \begin{subfigure}[t]{1\columnwidth}
    \centering
        \begin{tikzpicture} 
    \begin{axis}[%
    style={font=\footnotesize},
    hide axis,
    xmin=10,
    xmax=50,
    ymin=0,
    ymax=0.4,
    legend columns=3,
    column sep=1ex,
    legend image post style={scale=0.7}, 
    legend style={draw=none,nodes={anchor=east}},
    ]
    \addlegendimage{area legend, fill=red, draw=red, fill opacity=0.3, draw opacity=0.3}
    \addlegendentry{Naive};
    \addlegendimage{area legend, fill=blue, draw=blue, fill opacity=0.3, draw opacity=0.3}
    \addlegendentry{\stealing{}};
    \addlegendimage{dashed, line width=0.3mm};
    \addlegendentry{Average value};

    \end{axis}
\end{tikzpicture}
    \end{subfigure}

    \begin{subfigure}[t]{0.48\columnwidth}
    \centering
        \begin{tikzpicture}
\begin{axis}[
ybar,
style={font=\footnotesize},
xlabel={Semantic Similarity},
ylabel={Frequency (\%)},
width=1\linewidth,
scaled y ticks=false,
xtick pos=bottom,
ytick pos=left,
xtick style={draw=none},
xmin=0.45,
xmax=0.9,
ymin=0,
ymax=60,
clip=false,
height=3.3cm,
grid=both,
legend cell align=left,
legend columns=3,
legend style={
cells={align=left},
anchor=north,
at={(0.45,1.42)},
draw=none,
fill=none,
column sep=0.5ex,
},
legend image code/.code={%
    \draw[#1,draw] (0cm,-0.04cm) rectangle (0.12cm,0.12cm);
}
]
\addplot+[
    fill = red, draw=red, fill opacity=0.3, draw opacity=0.3,
    hist={bins=15,data min=0.6, data max=0.8},
    y filter/.expression={y*100/100},
] table[col sep=comma,y=naive_flux_diff] {fig/csv/attack2/prompt_steal_semantic_sim_dist.csv};
\addplot+[
    fill = blue, draw=blue, fill opacity=0.3, draw opacity=0.3,
    hist={bins=20,data min=0.45, data max=0.85},
    y filter/.expression={y*100/84},
] table[col sep=comma,y=ours_flux_diff] {fig/csv/attack2/prompt_steal_semantic_sim_dist.csv};


\node[draw=none, color=red, fill=none, inner sep=0.5mm] at (0.6,50) {0.67};
\draw[dashed, line width=0.3mm, color=red] (0.67,0) -- (0.67,60);
\node[draw=none, color=blue, fill=none, inner sep=0.5mm] at (0.82,50) {0.75};
\draw[dashed, line width=0.3mm, color=blue] (0.751,0) -- (0.751,60);

\end{axis}
\end{tikzpicture}
        \caption{FLUX with \diffusiondb{}}
        \label{fig:prompt-stealing-semantic-flux-diff}
    \end{subfigure}
    \hfill
    \begin{subfigure}[t]{0.48\columnwidth}
    \centering
        \begin{tikzpicture}
\begin{axis}[
ybar,
style={font=\footnotesize},
xlabel={Semantic Similarity},
ylabel={Frequency (\%)},
width=1\linewidth,
scaled y ticks=false,
xtick pos=bottom,
ytick pos=left,
xtick style={draw=none},
xmin=0.45,
xmax=0.9,
ymin=0,
ymax=60,
clip=false,
height=3.3cm,
grid=both,
legend cell align=left,
legend columns=3,
legend style={
cells={align=left},
anchor=north,
at={(0.45,1.42)},
draw=none,
fill=none,
column sep=0.5ex,
},
legend image code/.code={%
    \draw[#1,draw] (0cm,-0.04cm) rectangle (0.12cm,0.12cm);
}
]
\addplot+[
    fill = red, draw=red, fill opacity=0.3, draw opacity=0.3,
    hist={bins=15,data min=0.6, data max=0.8},
    y filter/.expression={y*100/100},
] table[col sep=comma,y=naive_flux_lex] {fig/csv/attack2/prompt_steal_semantic_sim_dist.csv};
\addplot+[
    fill = blue, draw=blue, fill opacity=0.3, draw opacity=0.3,
    hist={bins=20,data min=0.45, data max=0.85},
    y filter/.expression={y*100/96},
] table[col sep=comma,y=ours_flux_lex] {fig/csv/attack2/prompt_steal_semantic_sim_dist.csv};


\node[draw=none, color=red, fill=none, inner sep=0.5mm] at (0.6,50) {0.67};
\draw[dashed, line width=0.3mm, color=red] (0.67,0) -- (0.67,60);
\node[draw=none, color=blue, fill=none, inner sep=0.5mm] at (0.84,50) {0.78};
\draw[dashed, line width=0.3mm, color=blue] (0.78,0) -- (0.78,60);

\end{axis}
\end{tikzpicture}
        \caption{FLUX with \lexica{}}
        \label{fig:fig:prompt-stealing-semantic-flux-lexica}
    \end{subfigure}

    \begin{subfigure}[t]{0.48\columnwidth}
    \centering
        \begin{tikzpicture}
\begin{axis}[
ybar,
style={font=\footnotesize},
xlabel={Semantic Similarity},
ylabel={Frequency (\%)},
width=1\linewidth,
scaled y ticks=false,
xtick pos=bottom,
ytick pos=left,
xtick style={draw=none},
xmin=0.45,
xmax=0.9,
ymin=0,
ymax=60,
clip=false,
height=3.3cm,
grid=both,
legend cell align=left,
legend columns=3,
legend style={
cells={align=left},
anchor=north,
at={(0.45,1.42)},
draw=none,
fill=none,
column sep=0.5ex,
},
legend image code/.code={%
    \draw[#1,draw] (0cm,-0.04cm) rectangle (0.12cm,0.12cm);
}
]
\addplot+[
    fill = red, draw=red, fill opacity=0.3, draw opacity=0.3,
    hist={bins=15,data min=0.6, data max=0.8},
    y filter/.expression={y*100/100},
] table[col sep=comma,y=naive_sd3_diff] {fig/csv/attack2/prompt_steal_semantic_sim_dist.csv};
\addplot+[
    fill = blue, draw=blue, fill opacity=0.3, draw opacity=0.3,
    hist={bins=20,data min=0.45, data max=0.85},
    y filter/.expression={y*100/100},
] table[col sep=comma,y=ours_sd3_diff] {fig/csv/attack2/prompt_steal_semantic_sim_dist.csv};


\node[draw=none, color=red, fill=none, inner sep=0.5mm] at (0.6,50) {0.67};
\draw[dashed, line width=0.3mm, color=red] (0.67,0) -- (0.67,60);
\node[draw=none, color=blue, fill=none, inner sep=0.5mm] at (0.86,50) {0.81};
\draw[dashed, line width=0.3mm, color=blue] (0.805,0) -- (0.805,60);

\end{axis}
\end{tikzpicture}
        \caption{SD3 with \diffusiondb{}}
        \label{fig:fig:prompt-stealing-semantic-sd3-diff}
    \end{subfigure}
    \hfill
    \begin{subfigure}[t]{0.48\columnwidth}
    \centering
        \begin{tikzpicture}
\begin{axis}[
ybar,
style={font=\footnotesize},
xlabel={Semantic Similarity},
ylabel={Frequency (\%)},
width=1\linewidth,
scaled y ticks=false,
xtick pos=bottom,
ytick pos=left,
xtick style={draw=none},
xmin=0.45,
xmax=0.9,
ymin=0,
ymax=60,
clip=false,
height=3.3cm,
grid=both,
legend cell align=left,
legend columns=3,
legend style={
cells={align=left},
anchor=north,
at={(0.45,1.42)},
draw=none,
fill=none,
column sep=0.5ex,
},
legend image code/.code={%
    \draw[#1,draw] (0cm,-0.04cm) rectangle (0.12cm,0.12cm);
}
]
\addplot+[
    fill = red, draw=red, fill opacity=0.3, draw opacity=0.3,
    hist={bins=15,data min=0.6, data max=0.8},
    y filter/.expression={y*100/100},
] table[col sep=comma,y=naive_sd3_lex] {fig/csv/attack2/prompt_steal_semantic_sim_dist.csv};
\addplot+[
    fill = blue, draw=blue, fill opacity=0.3, draw opacity=0.3,
    hist={bins=20,data min=0.45, data max=0.85},
    y filter/.expression={y*100/100},
] table[col sep=comma,y=ours_sd3_lex] {fig/csv/attack2/prompt_steal_semantic_sim_dist.csv};


\node[draw=none, color=red, fill=none, inner sep=0.5mm] at (0.6,50) {0.67};
\draw[dashed, line width=0.3mm, color=red] (0.67,0) -- (0.67,60);
\node[draw=none, color=blue, fill=none, inner sep=0.5mm] at (0.85,50) {0.79};
\draw[dashed, line width=0.3mm, color=blue] (0.792,0) -- (0.792,60);

\end{axis}
\end{tikzpicture}
        \caption{SD3 with \lexica{}}
        \label{fig:fig:prompt-stealing-semantic-sd3-lexica}
    \end{subfigure}

    \caption{Semantic similarity distribution of recovered prompts ($n=100$). 
    }
    \label{fig:prompt-stealing-semantic}
\end{figure}

\begin{table}[t]
    \setlength{\tabcolsep}{4pt}
    \caption{Word-level prompt recovery quality.}
    \label{tab:word-level-success-rate}
    \centering
    \small
    \begin{tabular}{>{\centering\arraybackslash}m{1.7cm}cccccc}
        \toprule
        Dataset & Model & Method & BLEU & NERR & \makecell{BERT-\\Score} \\
        \midrule
    
        \multirow{4}{*}{\centering \diffusiondb{}} 
            & \multirow{2}{*}{FLUX} & Naive Probing  & 0.12 & 0.01 &0.83 \\
            &                       & \stealing{} & \textbf{0.22}&\textbf{0.17} &\textbf{0.85} \\
        \cmidrule{2-6}
            & \multirow{2}{*}{SD3} & Naive Probing & 0.12& 0.06& 0.80\\
            &                      & \stealing{} & \textbf{0.25} & \textbf{0.18} & \textbf{0.86}\\
        \midrule
           \multirow{4}{*}{\centering \lexica{}}  & \multirow{2}{*}{FLUX} & {Naive Probing}     & 0.11 & 0.01 &0.83 \\
            &                       & \stealing{} & \textbf{0.20}& \textbf{0.14}&\textbf{0.85} \\
        \cmidrule{2-6}
            & \multirow{2}{*}{SD3} & Naive Probing  & 0.11 & 0.02 & 0.83 \\
            &                      & \stealing{} &  \textbf{0.22} & \textbf{0.13} & \textbf{0.85} \\
        \bottomrule
    \end{tabular}
\end{table}

\begin{table}[t]
    \setlength{\tabcolsep}{4pt}
    \caption{Image recovery quality.}
    \label{tab:prompt-stealing-success-rate}
    \centering
    \small
    \begin{tabular}{>{\centering\arraybackslash}m{1.7cm}cccccc}
        \toprule
        Dataset & Model & Method & CLIP & PSNR & SSIM \\
        \midrule
    
        \multirow{4}{*}{\centering \diffusiondb{}} 
            & \multirow{2}{*}{FLUX} & Naive Probing     & 0.86 &20.89 &0.74 \\
            &                       & \stealing{}  & \textbf{0.90}&\textbf{24.95} &\textbf{0.82} \\
        \cmidrule{2-6}
            & \multirow{2}{*}{SD3} & Naive Probing     & 0.74& 15.02& 0.55\\
            &                      & \stealing{} & \textbf{0.85} & \textbf{19.92} & \textbf{0.75}\\
        \midrule
           \multirow{4}{*}{\centering \lexica{}}  & \multirow{2}{*}{FLUX} & Naive Probing     &0.88 & 20.11 &0.72 \\
            &                       & \stealing{} & \textbf{0.95}& \textbf{25.62}&\textbf{0.85} \\
        \cmidrule{2-6}
            & \multirow{2}{*}{SD3} & Naive Probing  & 0.75 & 15.39 & 0.531 \\
            &                      & \stealing{} &  \textbf{0.84} & \textbf{19.72} & \textbf{0.71} \\
        \bottomrule
    \end{tabular}
\end{table}

\subsubsection{Overall Success Rate}

We first conduct an experiment to show the prompt and image similarity on both models and both datasets, using the metrics in \Cref{subsec:stealing_setup}.
We process 100 rounds for each configuration and try to recover at least one prompt for each round. 

\noindent\textbf{Prompt semantic similarity.}
\Cref{fig:prompt-stealing-semantic} shows the distribution of the CLIP semantic similarity between the target prompts and the recovered prompts. 
On average, prompts recovered by \stealing{} achieve 0.75--0.81 similarity scores, much higher than the naive baseline which is around 0.67. 
According to prior studies, such high prompt similarity can be considered successful \cite{xinyue_prompt_stealing,cross_modal_prompt_inversion}. 
A few prompts recovered by \stealing{} has a similarity score lower than 0.65 due to false positives from our \textit{classifier} in those specific cases. We will discuss the impact of the false positives in \Cref{subsubsec:prompt-stealing-ablation-classifier}.

\noindent \textbf{Prompt word-level similarity.}
\Cref{tab:word-level-success-rate} displays the success rate of \stealing{} under word-level similarity metrics. 
For both BLEU and NERR, \stealing{} performs much higher accuracy scores than Naive Probing. 
The reason is that Naive Probing only considers the semantic similarity due to the approximate caching mechanism. 
In contrast, \stealing{} exploits a 124M-parameter GPT-2 model to summarize a cluster of probing prompts that hit the same cached prompt, achieving high accuracy.
While BERTScore’s consideration of synonyms narrows the margin, \stealing{} still consistently outperforms Naive Probing in BERTScore.

\noindent\textbf{Image similarity.}
\Cref{tab:prompt-stealing-success-rate} presents the image similarity compared to the target image.
\stealing{} outperforms the baseline in all three metrics. 
Specifically, the PSNR score can exceed 25 and achieve over 19 for both datasets and models, indicating a negligible difference compared to the target image \cite{fardo2016formalevaluationpsnrquality}.
For both the baseline and \stealing{}, FLUX has better similarity scores than SD3 across all metrics. 
This is because FLUX's early cached states contain clearer content than SD3, resulting in more similar output. 

\subsubsection{Number of Probing Prompts}
\label{subsubsec:prompt-stealing-cost}

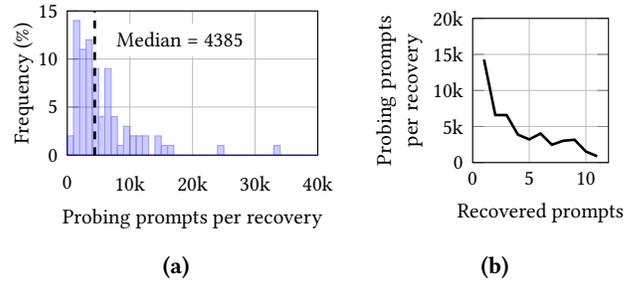
\begin{figure}
\begin{subfigure}[b]{0.58\linewidth}
    \centering
    \begin{tikzpicture}
\begin{axis}[
ybar,
style={font=\footnotesize},
xlabel={Probing prompts per recovery},
ylabel={Frequency (\%)},
ylabel shift={-5pt},
width=1\linewidth,
scaled y ticks=false,
xtick pos=bottom,
ytick pos=left,
scaled x ticks=false,
xtick = {0,10000,20000,30000,40000},
xticklabels={0, 10k, 20k, 30k, 40k},
xtick style={draw=none},
xmin=0,
xmax=40000,
ymin=0,
ymax=15,
clip=true,
height=3.5cm,
grid=both,
legend cell align=left,
legend columns=3,
legend style={
cells={align=left},
anchor=north,
at={(0.5,1.35)},
draw=none,
fill=none,
column sep=0.5ex,
},
legend image code/.code={%
    \draw[#1,draw] (0cm,-0.04cm) rectangle (0.12cm,0.12cm);
}
]
\addplot+[
    fill = blue, draw=blue, fill opacity=0.2, draw opacity=0.4,
    hist={bins=200,data min=0, data max=200000},
] table[col sep=comma,y=request per recovery] {fig/csv/attack2/prompt_steal_total_probe_requests.csv};

\draw [dashed, line width=0.35mm] (4385,0) -- (4385,15);
\node [draw=none,align=left,fill=white] at (18000,12) {Median = \SIx{4385}};

\end{axis}
\end{tikzpicture}
    \caption{\label{fig:num_probe_prompts}}
\end{subfigure}
\begin{subfigure}[b]{0.4\linewidth}
    \centering
    \begin{tikzpicture}
\begin{axis}[
    style={font=\footnotesize},
    scaled y ticks=false,
    ylabel shift={-3pt},
    bar width=6pt,
    clip=false,
    height=3.5cm,
    width=1\hsize,
    xmin=0,
    xmax=12,
    ymin=1,
    ymax=20000,
    ytick = {0,5000,10000,15000,20000},
    yticklabels={0,5k,10k,15k,20k},
    xlabel={Recovered prompts},
    ylabel style={align=center, text width=3cm},
    ylabel={Probing prompts per recovery},
    xtick style={draw=none},
    grid=both,
    legend columns=1,
    legend cell align=left,
    legend style={
        cells={align=left},
        anchor=south,
        at={(1.2,0.5)},
        draw=none,
        fill=none,
        column sep=0.5ex,
    },
    legend image code/.code={%
    \draw[#1, draw] (0cm,-0.05cm)
    rectangle (0.2cm,0.1cm);
    }
]

\addplot [black, mark=none, line width=0.35mm] table[col sep=comma, x=total_recovered, y=median_req] {fig/csv/attack2/prompt_steal_amortized_avg.csv};

\end{axis}

\end{tikzpicture}
\vspace{-3.5mm}
    \caption{\label{fig:amortized_cost}}
\end{subfigure}
    \caption{(a) Distribution of probing prompts and (b) amortized probing number over recovered prompts. }
\end{figure}

To steal a prompt, the attacker needs to send a large number of probing prompts before collecting a good number of hits for recovery. 
\Cref{fig:num_probe_prompts} shows the distribution of the number of probing prompts per recovery (stolen prompts).
Despite a long tail, the median is \SIx{4385}.
The high number is mainly due to the first stage of probing. 
After finding the initial hits, the \textit{prompt constructor} can generate more specific probing prompts in the second stage.
\Cref{fig:amortized_cost} shows the median number of probing prompts per recovery over different numbers of recovered prompts after each recovery execution. 
The number of probing prompts is amortized, as the number of recovered prompts in a round increases (as low as \SIx{874} when 11 prompts are recovered in a round).

\subsubsection{Ablation Study on Classifier}
\label{subsubsec:prompt-stealing-ablation-classifier}

\begin{figure}[t]
  \centering
  \begin{subfigure}[b]{1\linewidth}
    \centering
    \begin{tikzpicture} 
    \begin{axis}[%
    width=1\hsize,
    style={font=\footnotesize},
    hide axis,
    xmin=10,
    xmax=50,
    ymin=0,
    ymax=0.4,
    legend columns=3,
    column sep=1ex,
    legend image post style={scale=0.7}, 
    legend style={
        cells={align=center},
        anchor=south,
        at={(0.5,0.5)},
        draw=none,
        fill=none,
        column sep=0.5ex,
    },
    ]
    \addlegendimage{area legend, fill = red, draw=red, fill opacity=0.2, draw opacity=0.3}
    \addlegendentry{Greedy};
    \addlegendimage{area legend, fill = blue, draw=blue, fill opacity=0.2, draw opacity=0.3}
    \addlegendentry{\stealing{}};
    \addlegendimage{area legend, fill = green!35!black, densely dashed, line width=0.2mm, draw=green!35!black,draw opacity=0.7,fill opacity=0.1}
    \addlegendentry{Ideal};
    \end{axis}
\end{tikzpicture}
    \end{subfigure}
    
    \begin{subfigure}[b]{0.48\linewidth}
    \begin{tikzpicture}
\begin{axis}[
ybar,
style={font=\footnotesize},
xlabel={Semantic Similarity},
ylabel={Frequency (\%)},
width=1\linewidth,
scaled y ticks=false,
xtick pos=bottom,
ytick pos=left,
xtick style={draw=none},
xmin=0,
xmax=1,
ymin=0,
ymax=50,
clip=false,
height=3.3cm,
grid=both,
legend cell align=left,
legend columns=3,
legend style={
cells={align=left},
anchor=north,
at={(0.45,1.42)},
draw=none,
fill=none,
column sep=0.5ex,
},
legend image code/.code={%
    \draw[#1,draw] (0cm,-0.04cm) rectangle (0.12cm,0.12cm);
}
]
\addplot+[
    fill = red, draw=red, fill opacity=0.2, draw opacity=0.3,
    hist={bins=20,data min=0, data max=1},
    y filter/.expression={y*100/94},
] table[col sep=comma,y=greedy_diffdb_flux] {fig/csv/attack2/prompt_steal_classifier.csv};
\addplot+[
    fill = blue, draw=blue, fill opacity=0.2, draw opacity=0.3,
    hist={bins=20,data min=0, data max=1},
    y filter/.expression={y*100/100},
] table[col sep=comma,y=cache_exposer_diffdb_flux] {fig/csv/attack2/prompt_steal_classifier.csv};
\addplot+[
    fill = green!35!black, densely dashed, line width=0.2mm, draw=green!35!black,draw opacity=0.7,fill opacity=0.1,
    hist={bins=20,data min=0, data max=1},
    y filter/.expression={y*100/100},
] table[col sep=comma,y=ideal_diffdb_flux] {fig/csv/attack2/prompt_steal_classifier.csv};


\end{axis}
\end{tikzpicture}
    \caption{FLUX}
    \label{fig:prompt-stealing-effect-classifier-flux}
    \end{subfigure}
    \begin{subfigure}[b]{0.48\linewidth}
    \begin{tikzpicture}
\begin{axis}[
ybar,
style={font=\footnotesize},
xlabel={Semantic Similarity},
ylabel={Frequency (\%)},
width=1\linewidth,
scaled y ticks=false,
xtick pos=bottom,
ytick pos=left,
xtick style={draw=none},
xmin=0,
xmax=1,
ymin=0,
ymax=50,
clip=false,
height=3.3cm,
grid=both,
legend cell align=left,
legend columns=3,
legend style={
cells={align=left},
anchor=north,
at={(0.45,1.42)},
draw=none,
fill=none,
column sep=0.5ex,
},
legend image code/.code={%
    \draw[#1,draw] (0cm,-0.04cm) rectangle (0.12cm,0.12cm);
}
]
\addplot+[
    fill = red, draw=red, fill opacity=0.2, draw opacity=0.3,
    hist={bins=20,data min=0, data max=1},
    y filter/.expression={y*100/94},
] table[col sep=comma,y=greedy_diffdb_sd3] {fig/csv/attack2/prompt_steal_classifier.csv};
\addplot+[
    fill = blue, draw=blue, fill opacity=0.2, draw opacity=0.3,
    hist={bins=20,data min=0, data max=1},
    y filter/.expression={y*100/100},
] table[col sep=comma,y=cache_exposer_diffdb_sd3] {fig/csv/attack2/prompt_steal_classifier.csv};
\addplot+[
    fill = green!35!black, densely dashed, line width=0.2mm, draw=green!35!black,draw opacity=0.7,fill opacity=0.1,
    hist={bins=20,data min=0, data max=1},
    y filter/.expression={y*100/100},
] table[col sep=comma,y=ideal_diffdb_sd3] {fig/csv/attack2/prompt_steal_classifier.csv};



\end{axis}
\end{tikzpicture}
    \caption{SD3}
    \label{fig:prompt-stealing-effect-classifier-sd3}
    \end{subfigure}
  \caption{\label{fig:prompt-stealing-effect-classifier} The effectiveness of \textit{classifier} ($n=100$). }
\end{figure}
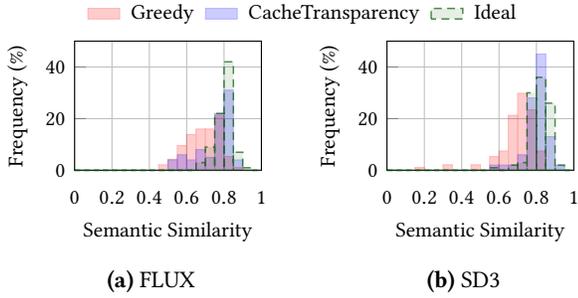

We evaluate the effectiveness of the \textit{classifier}.
Because the classifier takes both the prompt and the image as input to determine structural similarity, we evaluate both FLUX and SD3 using the main dataset, \diffusiondb{}.
We include two additional methods as comparison points:
(a) A \emph{greedy} method that uses all prompts that hit any cached prompts for recovery, without grouping them by their structural similarity. 
(b) An \emph{ideal} scenario that classifies the probing prompts 100\,\% correct. 
The prompt constructor generates the same probing prompts for \stealing{} and these methods.
\Cref{fig:prompt-stealing-effect-classifier} shows that \stealing{} is only 3.01\,\% lower than \emph{ideal} on average, following similar distributions. 
In comparison, without considering structural similarity, recovered prompts by the \textit{greedy} method have a lower similarity (0.69 on average).

\begin{figure}
    \centering
    \begin{tikzpicture}
\begin{axis}[
    style={font=\footnotesize},
    ybar,
    bar width=6pt,
    clip=false,
    height=3.3cm,
    width=0.8\hsize,
    xmin=-0.5,
    xmax=3.5,
    ymin=0.4,
    ymax=1,
    xtick = {0,1,2,3},
    xlabel = {Number of false positives in classifier},
    ylabel={Prompt similarity},
    xtick style={draw=none},
    ymajorgrids=true,
    legend columns=1,
    legend cell align=left,
    legend style={
        cells={align=left},
        anchor=south,
        at={(1.2,0.5)},
        draw=none,
        fill=none,
        column sep=0.5ex,
    },
    legend image code/.code={%
    \draw[#1, draw] (0cm,-0.05cm)
    rectangle (0.2cm,0.1cm);
    }
]

\addplot [draw=blue!60!white,fill=blue!40!white,error bars/.cd,y dir=both,y explicit] table[col sep=comma, x=FP, y=Similarity_FLUX, y error=Similarity_FLUX_err] {fig/csv/attack2/prompt_steal_fp.csv};

\addplot [draw=black!60!white,fill=black!20!white,error bars/.cd,y dir=both,y explicit] table[col sep=comma, x=FP, y=Similarity_SD3, y error=Similarity_SD3_err] {fig/csv/attack2/prompt_steal_fp.csv};
\legend{FLUX, SD3}

\end{axis}

\end{tikzpicture}
    \caption{Sensitivity to the false positive in classifier (prompts from \diffusiondb{}).}
    \label{fig:prompt-stealing-fp-num}
\end{figure}
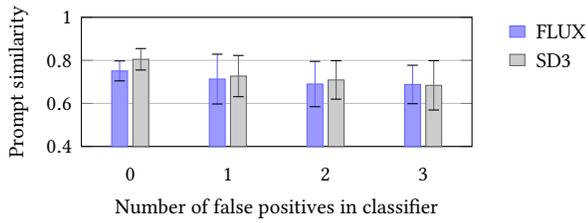

A false positive occurs when a probing prompt hits a different cached prompt but is mistakenly classified as hitting the target prompt in the cache.
In our experiment, SD3 and FLUX have \SIx{0.36} and \SIx{1.63} false positives on average.
We further evaluate its impact by randomly inserting variable numbers of false positives into the ideal scenario.
\Cref{fig:prompt-stealing-fp-num} shows that the similarity declines with more false positives.
The experiment also indicates that the similarity drops significantly with only 1 false positive, which highlights the importance of an accurate classifier. 

\subsubsection{Ablation Study on Prompt Recoverer}
\label{subsubsec:prompt-stealing-ablation-recoverer}

The prompt recoverer component converts embeddings back to prompts. We evaluate its effectiveness by measuring the similarity loss during the conversion. 
Because it does not take images as input, we only evaluate FLUX using \diffusiondb{}.
We compare the embedding of the target cached prompt directly with the \textit{embedding predictor}'s output (before recovery), as well as the embedding after conversion back to the prompt (after recovery). 
\Cref{fig:prompt-stealing-effect-prompt-recover} presents the distribution of similarity scores.
Compared to the embedding before recovery, \textit{prompt recoverer} only incurs 4.4\,\% similarity loss on average, preserving the similarity.

\begin{figure}
\begin{minipage}[b]{0.48\linewidth}
  \centering
  \begin{tikzpicture}
\begin{axis}[
ybar,
style={font=\footnotesize},
xlabel={Semantic Similarity},
ylabel={Frequency (\%)},
width=1\linewidth,
scaled y ticks=false,
xtick pos=bottom,
ytick pos=left,
xmin=0,
xmax=1,
ymin=0,
ymax=60,
clip=false,
height=3.3cm,
grid=both,
legend cell align=left,
legend columns=3,
legend style={
cells={align=left},
anchor=north,
at={(0.45,1.35)},
draw=none,
fill=none,
column sep=0.5ex,
},
legend image code/.code={%
    \draw[#1,draw] (0cm,-0.04cm) rectangle (0.12cm,0.12cm);
}
]
\addplot+[
    fill = red, draw=red, fill opacity=0.2, draw opacity=0.4,
    hist={bins=20,data min=0, data max=1},
    y filter/.expression={y*100/84},
] table[col sep=comma,y=diffdb_flux_before] {fig/csv/attack2/prompt_steal_recoverer.csv};
\addplot+[
    fill = blue, draw=blue, fill opacity=0.2, draw opacity=0.4,
    hist={bins=20,data min=0, data max=1},
    y filter/.expression={y*100/100},
] table[col sep=comma,y=diffdb_flux_after] {fig/csv/attack2/prompt_steal_recoverer.csv};
\legend{Before recovery, After recovery};

\end{axis}
\end{tikzpicture}
  \caption{\label{fig:prompt-stealing-effect-prompt-recover} The effectiveness of \textit{prompt recoverer} (FLUX with \diffusiondb{}, $n=100$). }
\end{minipage}
\hfill
\begin{minipage}[b]{0.48\linewidth}
    \centering
    \begin{tikzpicture}
\begin{axis}[
ybar,
style={font=\footnotesize},
xlabel={Semantic Similarity},
ylabel={Frequency (\%)},
width=1\linewidth,
scaled y ticks=false,
xtick pos=bottom,
ytick pos=left,
xmin=0.2,
xmax=1,
ymin=0,
ymax=60,
clip=false,
height=3.3cm,
grid=both,
legend cell align=left,
legend columns=3,
legend style={
cells={align=left},
anchor=north,
at={(0.45,1.35)},
draw=none,
fill=none,
column sep=0.5ex,
},
legend image code/.code={%
    \draw[#1,draw] (0cm,-0.04cm) rectangle (0.12cm,0.12cm);
}
]
\addplot+[
    fill = red, draw=red, fill opacity=0.3, draw opacity=0.4,
    hist={bins=30,data min=0.2, data max=1},
    y filter/.expression={y*100/84},
] table[col sep=comma,y=diffdb_flux_llama_semantic] {fig/csv/attack2/prompt_steal_llama.csv};
\addplot+[
    fill = blue, draw=blue, fill opacity=0.3, draw opacity=0.4,
    hist={bins=30,data min=0.2, data max=1},
    y filter/.expression={y*100/84},
] table[col sep=comma,y=diffdb_flux_ours_semantic] {fig/csv/attack2/prompt_steal_llama.csv};
\legend{Llama 3 70B, \stealing{}}

\end{axis}
\end{tikzpicture}
    \caption{\textit{Prompt recoverer} vs. Llama-3 70B (FLUX with \diffusiondb{}, $n=100$). 
    }
    \label{fig:prompt-stealing-effect-emb}
\end{minipage}
\end{figure}

    
    

We also compare our prompt recovery technique to an LLM-based method. 
We compare the similarity score of prompts recovered by the combination of \textit{Embedding Predictor} and \textit{Prompt Recover} in \stealing{} with those generated by a Llama-3~70B model \cite{llama3}.
As \Cref{fig:prompt-stealing-effect-emb} illustrated, \stealing{} achieves 10.4\% higher similarity scores than Llama.
It is worth noting that the recovery models in \stealing{} are lightweight compared to a 70B Llama-3, being able to execute on local devices (\eg{} a gaming GPU). This low hardware requirement further enhances the stealthiness of \stealing{}.

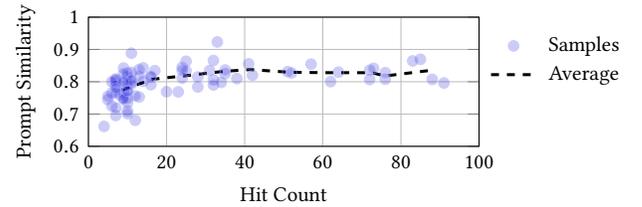
\begin{figure}
    \centering
    \begin{tikzpicture}
\begin{axis}[
    style={font=\footnotesize},
    clip=false,
    height=3.3cm,
    width=0.8\hsize,
    xmin=0,
    xmax=100,
    ymin=0.6,
    ymax=1,
    xlabel = {Hit Count},
    ylabel={Prompt Similarity},
    xtick style={draw=none},
    grid=major,
    legend columns=1,
    legend cell align=left,
    legend style={
        cells={align=left},
        anchor=south,
        at={(1.2,0.4)},
        draw=none,
        fill=none,
        column sep=0.5ex,
    },
]

\addplot+ [only marks, mark size=2pt, draw=blue!40!white, fill=blue!60!white, draw opacity=0.5, fill opacity=0.2] table[col sep=comma, x=hit count, y=semantic sim] {fig/csv/attack2/prompt_steal_hit_count_scatter.csv};

\addplot+ [draw=black, dashed, line width=0.4mm, no markers] table[col sep=comma, x=hit count, y=semantic sim] {fig/csv/attack2/prompt_steal_hit_count_avg.csv};

\legend{Samples, Average}

\end{axis}

\end{tikzpicture}
    \caption{Sensitivity to hit count (FLUX with \diffusiondb{}).}
    \label{fig:prompt-stealing-sens-hit-number}
\end{figure}

\subsubsection{Sensitivity Study on Hit Count}
We analyze the impact of hit count, a key factor in the prompt recovery process, using the primary model, FLUX, and the \diffusiondb{} dataset.
We evaluate 100 ideal experiments where the \textit{classifier} is 100\,\% correct and group the experiments by the number of prompts that hit the target cached prompt. 
\Cref{fig:prompt-stealing-sens-hit-number} illustrates the relationship between semantic similarity and the number of hits, where achieving a high similarity score (over 0.8) requires more than 17 hits.  
When the hit count increases over 26, the recovery similarity plateaus, indicating that the attacker can stop probing the image generation service once the classifier has collected 26 prompts that hit the same cached prompt.

\subsection{Discussions}
To the best of our knowledge, \stealing{} is the first prompt stealing attack for text-to-image diffusion models via the approximate cache. 
Compared to prior works \cite{xinyue_prompt_stealing,cross_modal_prompt_inversion,wu-etal-2025-vulnerability}, \stealing{} can steal prompts without requiring users to share their output image to the public. 
This enhances the capability of the prompt-stealing attack.
\stealing{} also constructs multiple probing prompts and exploits a timing classifier to determine the hit situation of those prompts, like cache side-channel attacks \cite{DeepCache,Peek-a-Walk,Spec-o-Scope,MeshUp,Yu2023M1,Liu_SP_2015_Last,Packet_Chasing,FLUSH-RELOAD,Kayaalp_DAC_2016_A_High}. 
Differently, this attack deals with similarity-based cache retrieval, and uses a \textit{structural classifier} and an \textit{embedding predictor} to recover the target prompt from a set of probing prompts. 

Like the covert channel (\Cref{subsec:covert_channel_discussion}), \stealing{} also works without cache timing. 
Without timing information, the \textit{structural classifier} can directly cluster prompts by the cache they hit based on the output images. 
However, this is more costly as the classifier needs to compare each pair of the probing prompts (14k). In contrast, the \textit{latency classifier} significantly narrows the search space.
\section{\poison{}: Image Poisoning Attack}
\label{sec:poison}

If a prompt hits the approximate cache, the output image resembles the cached one when it is similar with the cache prompt.
Therefore, attackers can inject prompts with their preferred information (e.g., logos) into the cached states, which will later be reused by victims. 
In this section, we present an image poisoning attack, \poison{}.

\subsection{Attack Model} 

\begin{figure}[t]
  \begin{center}
  \includegraphics[width=\linewidth]{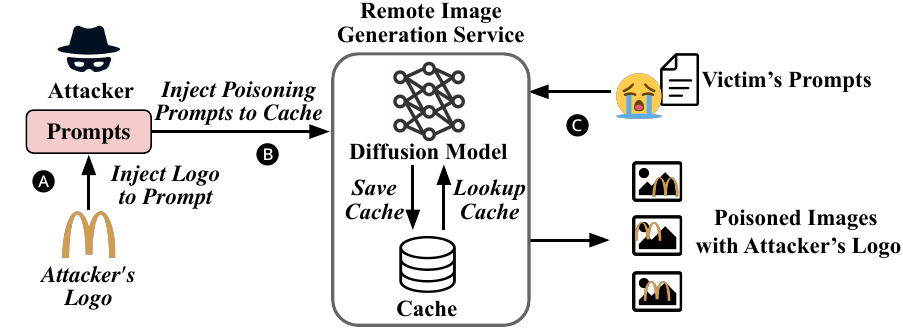}
  \end{center}
  \caption{\label{fig:poison_scenario} The setup and overview of \poison{}. }
\end{figure}

\Cref{fig:poison_scenario} depicts the setup of this attack.
We assume that an image generation service is the same as the previous covert channel (\Cref{subsec:covert_channel_attack_model}) and \stealing{} attack (\Cref{subsec:stealing_attack_mode}).
Likewise, we assume the attacker can only access the image generation service via prompts.
The attack's goal is to first inject prompts with information under their control and leave such information in user generations. 
The attacker has no prior knowledge about the existing caches in the cache system. Thus they have to steal the prompts first and then embed extra information into those prompts.
In this attack, we use logos to demonstrate the attacker's poisoning capabilities. The injected logos can be used for advertisement purposes.

\subsection{Attack Design}\label{subsec:poison_design}

\Cref{fig:poison_scenario} illustrates the overview of \poison{}. 
The attacker first injects the logo into prompts (\scalebox{0.7}{\circled{A}}) that were stolen using \stealing{} and then sends them to the approximate cache (\scalebox{0.7}{\circled{B}}). 
When a victim hits poisoned prompts in the approximate cache, the output will display the same logo (according to \Cref{box:primitive_2}), even if the victim did not mention it (\scalebox{0.7}{\circled{C}}). 
Because \stealing{} constructs probing prompts with commonly used words, the \poison{} attack that is based on these stolen prompts enables cache hits but does not guarantee the degree of popularity of the poisoned prompts. 
Our evaluation shows that \poison{} pollutes a substantial number of user prompts even though these prompts can be less popular (discussed in \Cref{subsubsec:poison-success}.)

\subsubsection{Challenge} 
To inject a cache that contains a logo, the logo's text description needs to be first included in the prompt and then cached. 
However, directly appending the logo to an arbitrary prompt is less likely to be hit by user prompts.

\begin{figure}[t]
  \begin{center}
  \includegraphics[width=\linewidth]{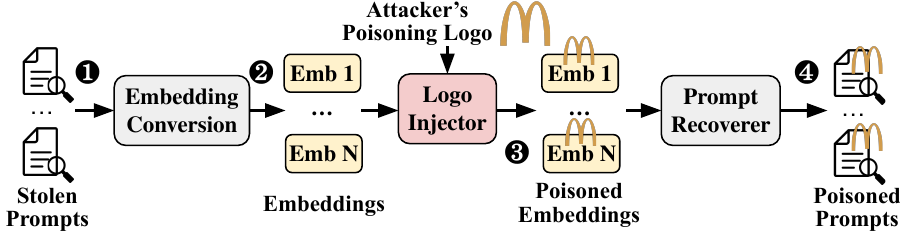}
  \end{center}
  \caption{\label{fig:poison_design} Design of \poison{}.}
\end{figure}

\subsubsection{Solution}
\Cref{fig:poison_design} depicts the design. 
First, the attacker steals cached prompts using the \stealing{} attack (\scalebox{0.8}{\circled{1}}).
These prompts are good representations of other user prompts and are more likely to be hit. 
Then, the attacker converts these prompts to embeddings (\scalebox{0.8}{\circled{2}}) and uses the \textit{logo injector} to embed logo descriptions into the embeddings (\scalebox{0.8}{\circled{3}}).  
Afterward, the attacker converts the embedding to a prompt with a similar model as the \textit{prompt recoverer} in \stealing{} (\scalebox{0.8}{\circled{4}}). 
This prompt contains the logo but remains similar to the stolen prompt. 

\noindent\textbf{Logo injector.} It is the core of this attack, which has two subcomponents: a model for \textit{logo insertion} and another for \textit{embedding prediction}.
The \textit{logo insertion} model is a small attention-based model that combines normal prompts with the attacker's logos. 
It takes the prompt embedding and the logo, and generates an embedding that contains both.
This is an effective way to insert logos, but the output may have low similarity with the original prompt, making it less likely to be hit by other users.
Therefore, we incorporate another \textit{embedding prediction model} to convert the injected prompts to be more similar to the original one.
The \textit{embedding prediction model} is similar to the one in \stealing{}, except that it includes the logo information.
It first generates an embedding $E$ and uses the \textit{logo insertion} model to guarantee that $E$ contains the logo information. We note the embedding with logo information as $E'$. 
To ensure that $E'$ preserves the semantics of both the original prompt and the logo, the model calculates two similarities: $S_{E'}$ for the similarity between $E'$ and the original prompt and $S_{\rm logo}$ for the similarity between $E'$ and the logo, and uses $(1 - S_{E'})^2 + (1 - S_{\rm logo})^2$ as the loss function.
$E'$ is optimized using SGD until the loss is below~$10^{-4}$.

\begin{figure}
  \begin{center}
  \includegraphics[width=\linewidth]{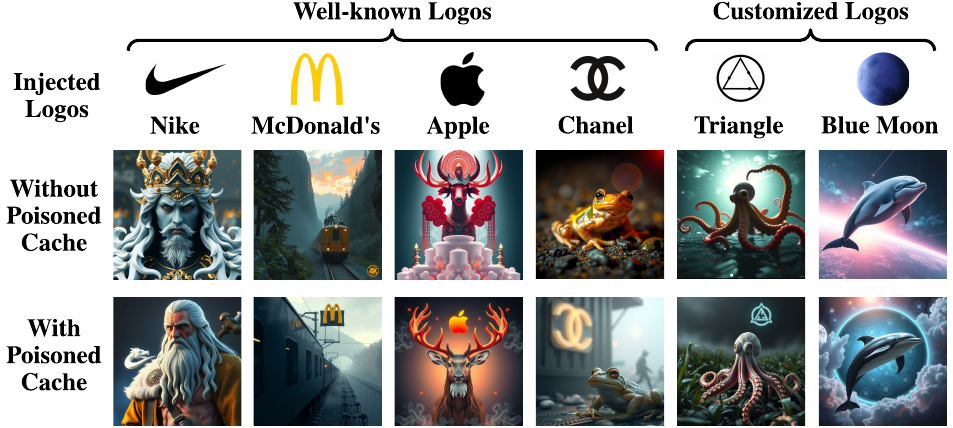}
  \end{center}
  \caption{\label{fig:poison_attack_example} Examples of \poison{} (generated by FLUX).}
\end{figure}

\subsection{Attack Setup}\label{subsec:poison_setup}

\textbf{Platform.} We use the same system in \Cref{subsubsec:cache_timing}.

\noindent\textbf{Model and Dataset. }
Similar to the Convert Channel, we also set FLUX and \diffusiondb{} as the main model and dataset, and SD3 and \lexica{} as an extension.
To evaluate the poisoning success rate, we send 1M prompts in \diffusiondb{}-Large by their timestamps and the whole \lexica{} randomly by its default order.

\noindent\textbf{Logos.}
We evaluate this attack with 6 logos, including 4 well-known logos (``Nike'', ``McDonald's'', ``Apple'', ``Chanel'') and 2 customized logos (``Triangle'', ``Blue Moon''), similar to the methodology in prior work \cite{jang2025silentbranding}.
We choose 50 target prompts stolen by \stealing{}, as described in \Cref{subsec:poison_design}. 
Thus, each logo has 50 poisoned prompts.

\noindent\textbf{\poison{} Training. }
The \textit{logo injector} model contains a 2-layer MLP and an attention layer.
To train this model, we use a training set based on 10\,\% of \diffusiondb{}-2M,
We randomly insert the logo descriptions into these prompts. 
The training objective is to maximize the cosine similarity between the model’s predicted embedding and the embedding of logo-inserted prompts.
To convert the poisoned embedding back to text, we exploit the same model as the \textit{prompt recoverer model} in \stealing{}. We retrain it with the new dataset with the added logo descriptions to ensure correct generation of the logo text.

\noindent\textbf{Baseline. }
We take the naive implementation of the poison attack that randomly injects the logo text into the stolen prompts and inserts them into the cache system as a baseline. 

\noindent\textbf{Metrics. }
We use the following metrics to evaluate this attack:

\begin{itemize}[leftmargin=*,noitemsep,partopsep=0pt,topsep=0pt,parsep=0pt]

  \item \textbf{Hit number} indicates the number of user prompts that hit the attack's poisoned prompts in the approximate cache.
  
  \item \textbf{Render rate} is the rate of user prompts that hit the cached prompts and contain a visible logo in the output images. The logo is not always rendered successfully by diffusion models, especially when the number of skipped steps is too small or the logo is complicated.
\end{itemize}

\begin{table}
  \centering
    \setlength{\tabcolsep}{1pt}
    \centering
    \small
    \caption{\label{tab:poison-attack-hit-render}Poison effectiveness.}
    \begin{tabular}{ccccccccc}
      \toprule
       \multirow{3}{*}{Method} & \multicolumn{4}{c}{FLUX} & \multicolumn{4}{c}{SD3} \\
          & \multicolumn{2}{c}{\diffusiondb{}} & \multicolumn{2}{c}{\lexica{}} & \multicolumn{2}{c}{\diffusiondb{}} & \multicolumn{2}{c}{\lexica{}} \\
        \cmidrule(lr){2-9}
        & Hit & Render & Hit & Render & Hit & Render & Hit & Render \\
        \midrule
        Direct Inject
            & \SIx{2624} & \SIx{1580} & \SIx{1059} & \SIx{744} & \SIx{3408} & \SIx{883} & \SIx{632} & \SIx{139} \\
        \poison{}
            & \textbf{3646}  & \textbf{2104} & \textbf{1226} & \textbf{863} & \textbf{4931} & \textbf{1360} & \textbf{838} & \textbf{180} \\
        \bottomrule
    \end{tabular}
  \hfill
\end{table}

\subsection{Results}

\Cref{fig:poison_attack_example} is an example of the poisoned output generated by FLUX. The first row presents the logos, and the second and third rows show the generated images with and without hitting the poisoned prompts.
When hitting a poisoned prompt, the logo embeds into the generated image.

\subsubsection{Poison Success Rate}
\label{subsubsec:poison-success}

We first evaluate the hit number and render success rate. 
We simulate user prompts using datasets as described in \Cref{subsec:poison_setup}.
In parallel, the attacker sends poisoned prompts periodically to determine whether the prompt was successfully inserted into the approximate cache based on the generation latency.
If a user prompt hits any poisoned cached state, we save the generated images and detect whether it contains the logo. The detection model is DINOv2 \cite{oquab2023dinov2}, similar to the method in prior work~\cite{jang2025silentbranding}.

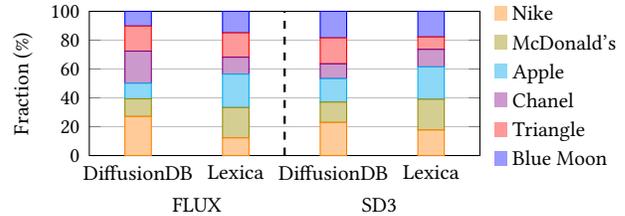
\begin{figure}
    \centering
    \begin{tikzpicture}
\begin{axis}[
    style={font=\footnotesize},
    ybar stacked,
    bar width=10pt,
    clip=true,
    height=3.5cm,
    width=0.8\hsize,
    xmin=-0.5,
    xmax=3.5,
    ymin=0,
    ymax=100,
    xtick = {0,1,2,3},
    x tick label style={align=center, text width=1.5cm}, 
    xticklabels = {\diffusiondb{},  \lexica{}, \diffusiondb{}, \lexica{}},
    xlabel={FLUX\hspace{0.22\linewidth}SD3},
    ylabel={Fraction (\%)},
    xtick style={draw=none},
    ymajorgrids=true,
    legend columns=1,
    legend cell align=left,
    legend style={
        inner sep=0pt,
        cells={align=left},
        anchor=north,
        at={(1.2,1.08)},
        draw=none,
        column sep=0ex,
    },
    legend image code/.code={%
        \draw[#1,draw=none] (0cm,-0.07cm) rectangle (0.17cm,0.12cm);
    }
]

\addplot+ [fill=orange!40!white, draw=orange] table[col sep=comma, y=Blue Moon, meta=model, x expr=\coordindex] {fig/csv/attack3/logo_hit_breakdown.csv};
\addplot+ [fill=olive!40!white, draw=olive] table[col sep=comma, y=Triangle, meta=model, x expr=\coordindex] {fig/csv/attack3/logo_hit_breakdown.csv};
\addplot+ [fill=cyan!40!white, draw=cyan] table[col sep=comma, y=Chanel, meta=model, x expr=\coordindex] {fig/csv/attack3/logo_hit_breakdown.csv};
\addplot+ [fill=violet!40!white, draw=violet] table[col sep=comma, y=Apple, meta=model, x expr=\coordindex] {fig/csv/attack3/logo_hit_breakdown.csv};
\addplot+ [fill=red!40!white, draw=red] table[col sep=comma, y=Mcdonald, meta=model, x expr=\coordindex] {fig/csv/attack3/logo_hit_breakdown.csv};
\addplot+ [fill=blue!40!white, draw=blue] table[col sep=comma, y=Nike, meta=model, x expr=\coordindex] {fig/csv/attack3/logo_hit_breakdown.csv};

\legend{Nike,McDonald's,Apple,Chanel,Triangle,Blue Moon}

\draw [dashed, black, line width=0.3mm] (1.5,0) -- (1.5,100);

\end{axis}

\end{tikzpicture}
    \caption{Breakdown of hits by logos.}
    \label{fig:poison-attack-logo-hit-breakdown}
\end{figure}

\begin{figure}[t]
\begin{subfigure}[t]{1\linewidth}
    \centering
    \begin{tikzpicture} 
    \begin{axis}[%
    width=1\hsize,
    style={font=\footnotesize},
    hide axis,
    xmin=10,
    xmax=50,
    ymin=0,
    ymax=0.4,
    legend columns=3,
    column sep=1ex,
    legend image post style={scale=0.7}, 
    legend style={
        cells={align=center},
        anchor=south,
        at={(0.5,0.5)},
        draw=none,
        fill=none,
        column sep=0.5ex,
    },
    ]
    \addlegendimage{area legend, draw=blue!60!white,fill=blue!40!white}
    \addlegendentry{\diffusiondb{}};
    \addlegendimage{area legend, draw=black!60!white,fill=black!20!white}
    \addlegendentry{\lexica{}};
    \end{axis}
\end{tikzpicture}
\end{subfigure}
\begin{subfigure}[b]{0.48\linewidth}
  \centering
  \begin{tikzpicture}
\begin{axis}[
    style={font=\footnotesize},
    ybar,
    bar width=3.5pt,
    clip=false,
    height=3.3cm,
    width=1\hsize,
    xmin=-0.5,
    xmax=5.5,
    ymin=0,
    ymax=100,
    xtick = {0,1,2,3,4,5},
    xticklabels={Nike, McDonald's, Apple, Chanel, Triangle, Blue Moon},
    xticklabel style={rotate=45,anchor=east},
    ylabel={Render Rate (\%)},
    xtick style={draw=none},
    ymajorgrids=true,
    legend columns=1,
    legend cell align=left,
    legend style={
        cells={align=left},
        anchor=south,
        at={(1.2,0.5)},
        draw=none,
        fill=none,
        column sep=0.5ex,
    },
    legend image code/.code={%
    \draw[#1, draw] (0cm,-0.05cm)
    rectangle (0.2cm,0.1cm);
    }
]

\addplot [draw=blue!60!white,fill=blue!40!white,  y filter/.expression={y*100},bar shift=-2pt] table[col sep=comma, meta=logo_name, x expr=\coordindex, y=flux_diffusiondb_rate] {fig/csv/attack3/render_rate.csv};
\addplot [draw=black!60!white,fill=black!20!white,  y filter/.expression={y*100},bar shift=2pt] table[col sep=comma, meta=logo_name, x expr=\coordindex, y=flux_lexica_rate] {fig/csv/attack3/render_rate.csv};

\end{axis}

\end{tikzpicture}
  \caption{FLUX} \label{fig:poison-attack-render-rate-flux}
\end{subfigure}
\begin{subfigure}[b]{0.48\linewidth}
  \centering
  \begin{tikzpicture}
\begin{axis}[
    style={font=\footnotesize},
    ybar,
    bar width=3.5pt,
    clip=false,
    height=3.3cm,
    width=1\hsize,
    xmin=-0.5,
    xmax=5.5,
    ymin=0,
    ymax=100,
    xtick = {0,1,2,3,4,5},
    xticklabels={Nike, McDonald's, Apple, Chanel, Triangle, Blue Moon},
    xticklabel style={rotate=45,anchor=east},
    ylabel={Render Rate (\%)},
    xtick style={draw=none},
    ymajorgrids=true,
    legend columns=1,
    legend cell align=left,
    legend style={
        cells={align=left},
        anchor=south,
        at={(1.2,0.5)},
        draw=none,
        fill=none,
        column sep=0.5ex,
    },
    legend image code/.code={%
    \draw[#1, draw] (0cm,-0.05cm)
    rectangle (0.2cm,0.1cm);
    }
]

\addplot [draw=blue!60!white,fill=blue!40!white,  y filter/.expression={y*100},bar shift=-2pt] table[col sep=comma, meta=logo_name, x expr=\coordindex, y=sd3_diffusiondb_rate] {fig/csv/attack3/render_rate.csv};
\addplot [draw=black!60!white,fill=black!20!white,  y filter/.expression={y*100},bar shift=2pt] table[col sep=comma, meta=logo_name, x expr=\coordindex, y=sd3_lexica_rate] {fig/csv/attack3/render_rate.csv};


\end{axis}

\end{tikzpicture}
  \caption{SD3} \label{fig:poison-attack-render-rate-sd3}
\end{subfigure}
  \caption{\label{fig:poison-attack-render-rate} Render success rate for different models.}
\end{figure}
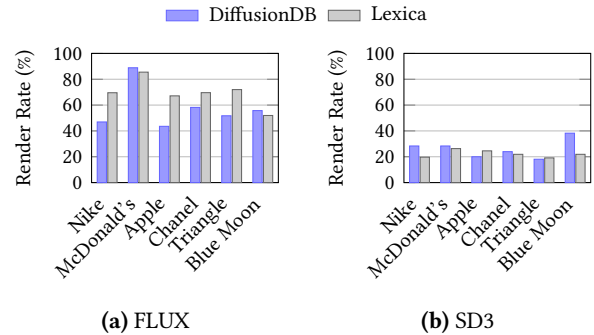

As \Cref{tab:poison-attack-hit-render} illustrates, \poison{} receives more hits than the direct injection baseline in all scenarios, due to the effective \textit{logo injector}.
Consequently, the success rate of \poison{} is also higher. 
\Cref{fig:poison-attack-logo-hit-breakdown} details the breakdown of hits by logos.
Because we use the same target prompts for all 6 logos, the variation among logos indicates how well a logo can be blended into the prompt.  
All 6 logos contribute to the total hit, indicating the generalization of this attack.

\begin{figure}
  \begin{center}
  \includegraphics[width=\linewidth]{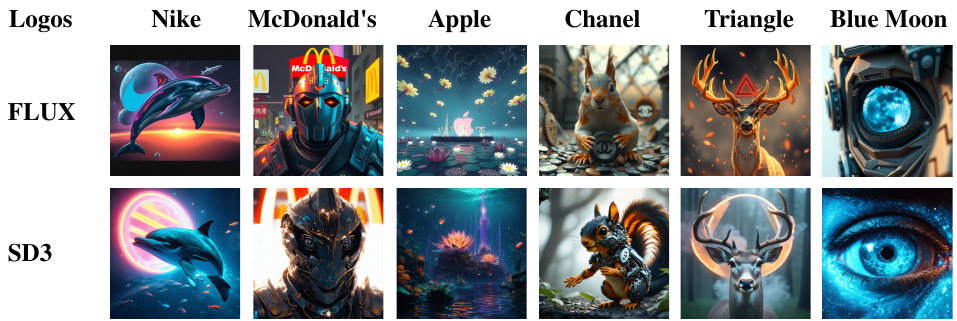}
  \end{center}
  \caption{\label{fig:sd3_vs_flux} Generation comparison between FLUX and SD3 (Median) when the logo is directly embedded into the prompt. }
\end{figure}

We also explore the render success rates among 6 logos. \Cref{fig:poison-attack-render-rate} demonstrates that FLUX achieves a higher success rate.
In contrast, SD3 struggles to preserve such visual content when reusing the cached states. 
The main reason is that SD3 (the median version) is less powerful than FLUX, resulting in weaker text–image alignment.
Even when the logo text is explicitly included in the prompt, SD3 often fails to generate the corresponding logo.
\Cref{fig:sd3_vs_flux} illustrates that, when a logo is embedded directly into the prompt, FLUX produces a clear logo, whereas SD3 does not.
Logo complexity also significantly impacts the success rate. For instance, \Cref{fig:poison-attack-render-rate-sd3} shows that ``Triangle'' exhibits the lowest success rate with SD3, as its complicated lines can vanish after multiple denoising steps, while ``Blue Moon'' achieves a high success rate due to its simpler structure.

\subsubsection{Sensitivity Study}
\begin{figure}[t]
    \centering
\begin{subfigure}[b]{0.48\linewidth}
    \begin{tikzpicture}
\begin{axis}[
    style={font=\footnotesize},
    ybar,
    ymode=log,
    bar width=6pt,
    clip=false,
    height=3.3cm,
    width=1\hsize,
    xmin=-0.3,
    xmax=2.3,
    ymin=1,
    ymax=100000,
    xtick = {0,1,2},
    xticklabels={\SI{1}{\giga\byte},\SI{10}{\giga\byte},\SI{100}{\giga\byte} },
    ylabel={Hit Count},
    xtick style={draw=none},
    ymajorgrids=true,
    legend columns=1,
    legend cell align=left,
    legend style={
        cells={align=left},
        anchor=south,
        at={(1.2,0.5)},
        draw=none,
        fill=none,
        column sep=0.5ex,
    },
    legend image code/.code={%
    \draw[#1, draw] (0cm,-0.05cm)
    rectangle (0.2cm,0.1cm);
    }
]

\addplot [draw=blue!60!white,fill=blue!40!white] table[col sep=comma, meta=size, x expr=\coordindex, y=count] {fig/csv/attack3/cache_size.csv};

\end{axis}

\end{tikzpicture}
    \caption{}\label{fig:poisoning_cache_size}
\end{subfigure}
\hfill
\begin{subfigure}[b]{0.48\linewidth}
    \begin{tikzpicture}
\begin{axis}[
    style={font=\footnotesize},
    ybar,
    ymode=log,
    bar width=6pt,
    clip=false,
    height=3.3cm,
    width=1\hsize,
    xmin=-0.3,
    xmax=2.3,
    ymin=1,
    ymax=100000,
    xtick = {0,1,2},
    xticklabels={FIFO, LRU, LCBFU},
    ylabel={Hit Count},
    xtick style={draw=none},
    ymajorgrids=true,
    legend columns=1,
    legend cell align=left,
    legend style={
        cells={align=left},
        anchor=south,
        at={(1.2,0.5)},
        draw=none,
        fill=none,
        column sep=0.5ex,
    },
    legend image code/.code={%
    \draw[#1, draw] (0cm,-0.05cm)
    rectangle (0.2cm,0.1cm);
    }
]

\addplot [draw=blue!60!white,fill=blue!40!white] table[col sep=comma, meta=policy, x expr=\coordindex, y=count] {fig/csv/attack3/replacement_policy.csv};

\end{axis}

\end{tikzpicture}
    \caption{}\label{fig:poisoning_replacement_policy}
\end{subfigure}
\caption{Poison effectiveness under (a) different cache sizes (LCBFU) and (b) replacement policies (\SI{100}{\giga\byte} cache).}
\end{figure}
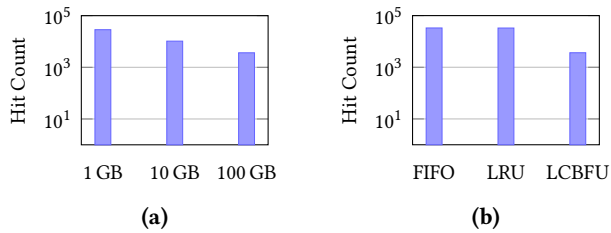

We finally explore \poison{}'s effectiveness under different cache configurations, like those studied in \Cref{subsubsec:covert_lifetime}. 
\Cref{fig:poisoning_cache_size} shows that the hit count is higher when the cache size is smaller, as the cached prompt that the poisoned one will replace gets evicted earlier under a smaller cache, allowing more user prompts to hit the attacker's poisoned cache. 
With the same cache size, \Cref{fig:poisoning_replacement_policy} shows the hit count is higher under FIFO and LRU than LCBFU because the target prompts in these schemes get evicted from the cache earlier. 
In LCBFU, cached prompts that were frequently hit in the past are harder to evict, leading to a lower hit count.

\subsection{Discussion}
Unlike previous poison attacks that pollute the training dataset \cite{jang2025silentbranding,wang2024the,guo2025rededitingrelationshipdrivenprecisebackdoor,Nightshade,pan24nipsfrom}, \poison{} poisons a serving system through prompts, without accessing or manipulating the training data. 
Therefore, this attack assumes a more realistic environment.
Moreover, \poison{} can work in combination with other poisoning attacks, further increasing the render success rate of the attacker's content.

\poison{} can also operate without the timing channel, which is only used for injecting poisoned prompts.
Without it, the attacker can analyze the generation to determine whether their poisoned prompt was processed from scratch or a cached prompt, \eg{} comparing it with the output from a separate system running the same model.
\section{Potential Defense Mechanisms}
\textbf{Random cache selection.} 
The approximate caching selects the cache of the highest similarity with the incoming prompt~\cite{nirvana, xia2025modmefficientservingimage}, which allows the sender's prompt with the same special word to be chosen.
The defense mechanism can randomly choose from a number of potential cached prompts. 
This randomization makes it significantly harder for a receiver to hit a specific entry, significantly decreasing the success rate of the covert channel.
Likewise, this defense also makes \stealing{} more costly, as more probing prompts are needed to collect the same number of hits for each target. 
For the poisoning attack, fewer prompts will hit the poisoned prompt in the approximate cache, thus reducing the spread of the attacker's content.

\noindent\textbf{Content filter.}
Similar to filters in generative models~\cite{Wang_CCS_2024_Moderator,Li_CCS_2024_SafeGen,sun2025pretender,zhangusd,villaexposing} that defend against not-suitable-for-work (NSFW) content, filters can also be specific to removing the attacker's content from the approximate cache. 
For example, to prevent the poisoning attack in this work, the service provider can incorporate a logo filter that detects logo-like objects in the image and potential logo descriptions in the prompt, to make sure both the cached prompts and the generated images do not contain logos.

\noindent\textbf{Malicious behavior detector.}
Stealing prompts from an approximate cache takes a large number of probing prompts as shown in \Cref{subsubsec:prompt-stealing-cost}.
A malicious behavior detector can monitor user's activities \cite{CloudRadar,NIGHTs-WATCH,SCADET,CHIAPPETTA20161162,Mushtaq2018mlsec,joshi2025hybriddeeplearningmodel}. 
When a user appears to abuse the generation system, such as flooding the system with prompts, the service provider can slow down or disable the user's generation.

\section{Conclusions}

Approximate caching improves the performance of diffusion models, but it also introduces significant security risks. We demonstrated three such attacks: a covert channel that transmits messages through the approximate cache, a prompt-stealing attack that reveals cached prompts, and a poisoning attack that embeds the attacker's content into user's generations.
These findings demonstrate that approximate caching introduces new security vulnerabilities which must be addressed before widespread adoption.


\section*{Acknowledgment}

We thank the anonymous reviewers and the shepherd for their valuable feedback.
This work is supported by a Discovery Grant from the Natural Sciences and Engineering Research Council of Canada (NSERC).

\appendix
\section*{Ethical Considerations}
\textbf{Decison to Conduct the Research.} Text-to-image diffusion models have been widely studied in academia and extensively used for creative generation in the industry. To alleviate the heavy computation overhead, approximate caching has been proposed and widely studied. There are many industrial deployment guides and academic studies about using approximate caching \cite{xia2025modmefficientservingimage,nirvana,bang-2023-gptcache,sun2024flexcacheflexibleapproximatecache}.
The goal of this work is to identify potential risks early and alert the diffusion community and startups, as some startups may simply follow these guides and open-source code, overlooking the security risks.

\textbf{Stakeholders.}
The attacks we identified pose potential threats to image generation service providers and their users, and prompt providers who design high-quality prompts.

\textbf{Impacts.}
Our paper identifies the potential risks in approximate caching, especially when such systems are deployed in production for public use:
First, an attacker may use the image generation service as a secret channel to transmit secret (and potentially illegal) messages.
Second, the prompt provider’s well-designed prompts may be stolen by the attackers.
Finally, service users may receive images with undesired patterns/logos, potentially containing attacker-controlled malicious information, in their generated images.

\textbf{Mitigation.}
We propose three methods to mitigate the vulnerabilities of caching for diffusion models. First, we propose random cache selection techniques to decrease the possibility that the receiver hits the sender's cached prompt, making approximate cache infeasible as a covert channel. Second, we propose to monitor the user behaviors to detect the potential malicious attackers in advance \cite{CloudRadar,NIGHTs-WATCH,SCADET,CHIAPPETTA20161162,Mushtaq2018mlsec,joshi2025hybriddeeplearningmodel}. This method identifies the probing operation in advance, protecting well-designed prompts by limiting the number of probing prompts, which leads to low-quality recovery for attackers. Finally, we propose to use content filters as in previous studies to filter the output images with unexpected logos \cite{Wang_CCS_2024_Moderator,Li_CCS_2024_SafeGen,sun2025pretender,zhangusd,villaexposing}. By doing so, the normal users will not receive images with unexpected content in them.
We have informed the authors and developers of NIRVANA, our reference system, about the risks and mitigation solutions.

\textbf{Attacks on Real Products.}
We conducted our experiments in our test systems, as our attacks aim to steal users' prompts and pollute the cache system of diffusion service. No real-world users are affected by our experiences. The test system, however, strictly follows the same assumptions and implementation of existing publications \cite{nirvana,sun2024flexcacheflexibleapproximatecache,xia2025modmefficientservingimage}, especially NIRVANA, the caching system developed by Adobe.

\section*{Open Science}
Our study adheres to open science principles. 
We open the code for our implementation of NIRVANA approximate cache from Adobe \cite{nirvana}, as well as code for all three attacks at the following link: \url{https://doi.org/10.5281/zenodo.17957900}. 
This repository also contains necessary scripts to reproduce the key results. 


\bibliographystyle{plain}
\bibliography{bib/ml, bib/sec, bib/misc, bib/sys}
\end{document}